\PassOptionsToPackage{table}{xcolor}
\documentclass[sigconf]{acmart}
\AtBeginDocument{%
  }

\copyrightyear{2027}
\acmYear{2027}
\setcopyright{cc}
\setcctype{by}
\acmConference[WSDM '27]{Proceedings of the Twentieth ACM International Conference on Web Search and Data Mining}{February 15--19, 2027}{Hong Kong, Hong Kong}
\acmBooktitle{Proceedings of the Twentieth ACM International Conference on Web Search and Data Mining (WSDM '27), February 15--19, 2027, Hong Kong, Hong Kong}
\acmDOI{10.1145/3862322.3863428}
\acmISBN{979-8-4007-3117-4/2027/02}
\usepackage{multirow}      
\usepackage{makecell}      
\usepackage{subcaption}
\usepackage{enumitem}

\begin{document}

\title[Semantic Trimming and Auxiliary Multi-step Prediction for Generative Recommendation]{Semantic Trimming and Auxiliary Multi-step Prediction for Generative Recommendation}


\author{Tianyu Zhan}
\authornote{Both authors contributed equally to this research.}
\affiliation{%
  \institution{Zhejiang University}
  \city{Hangzhou}
  \state{Zhejiang}
  \country{China}
}
\email{yuzt@zju.edu.cn}

\author{Kairui Fu}
\authornotemark[2]
\affiliation{%
  \institution{Zhejiang University}
  \city{Hangzhou}
  \state{Zhejiang}
  \country{China}
}
\email{fukairui.fkr@zju.edu.cn}

\author{Chengfei Lv}
\affiliation{%
  \institution{Alibaba}
  \city{Hangzhou}
  \state{Zhejiang}
  \country{China}
}
\email{chengfei.lcf@alibaba-inc.com}

\author{Zheqi Lv}
\authornotemark[3]
\affiliation{%
  \institution{Zhejiang University}
  \city{Hangzhou}
  \state{Zhejiang}
  \country{China}
}
\email{zheqilv@zju.edu.cn}

\author{Shengyu Zhang}
\authornote{Corresponding authors.}
\affiliation{%
  \institution{Zhejiang University}
  \city{Hangzhou}
  \state{Zhejiang}
  \country{China}
}
\affiliation{%
  \institution{Shanghai Institute for Advanced Study, Zhejiang University}
  \city{Shanghai}
  \country{China}
}
\email{sy\_zhang@zju.edu.cn}


\begin{abstract}
Generative Recommendation (GR) has recently transitioned from atomic item-indexing to Semantic ID (SID)-based frameworks to capture intrinsic item relationships and enhance generalization. However, the adoption of high-granularity SIDs leads to two critical challenges: prohibitive training overhead due to sequence expansion and unstable performance reliability characterized by non-monotonic accuracy fluctuations. We hypothesize that these two issues share a common source, which we term the Semantic Dilution Effect, where redundant tokens consume massive computation and dilute the already sparse learning signals in recommendation.
To counteract this, we propose STAMP (Semantic Trimming and Auxiliary Multi-step Prediction), a dual-end framework addressing low input information density and sparse output supervision. On the input side, Semantic Adaptive Pruning (SAP) dynamically filters redundancy during the forward pass, converting noise-laden sequences into compact, information-rich representations. On the output side, Multi-step Auxiliary Prediction (MAP) employs a multi-token objective to densify feedback, strengthening long-range dependency capture and ensuring robust learning signals despite compressed inputs. Coupling input purification with signal amplification improves training efficiency and representation capability. Experiments on public Amazon and large-scale industrial datasets show that, unlike more aggressive compression baselines that trade substantial accuracy for maximal speedup, STAMP achieves 1.23--1.38$\times$ speedup and 17.2\%--54.7\% VRAM reduction while maintaining or improving performance across multiple architectures.

\end{abstract}




\maketitle

\section{Introduction}

Generative Recommendation (GR) reformulates recommendation as end-to-end sequence generation~\cite{tiger,gr_1,gr_2,gr_3}. However, early frameworks using atomic item identifiers were limited by independent IDs, which hinder information sharing and often cause vocabulary explosion and cold-start problems for new items. Semantic Identifier (SID)-based Generative Recommendation (SID-GR)~\cite{tiger, onerec,forge, rw_15_22} addresses these issues by decomposing each item ID into a sequence of SIDs. The shared SID vocabulary enhances generalization across items (e.g., cross-domain recommendation), attracting recent attention.

\begin{figure}
    \centering
    \includegraphics[width=\linewidth]{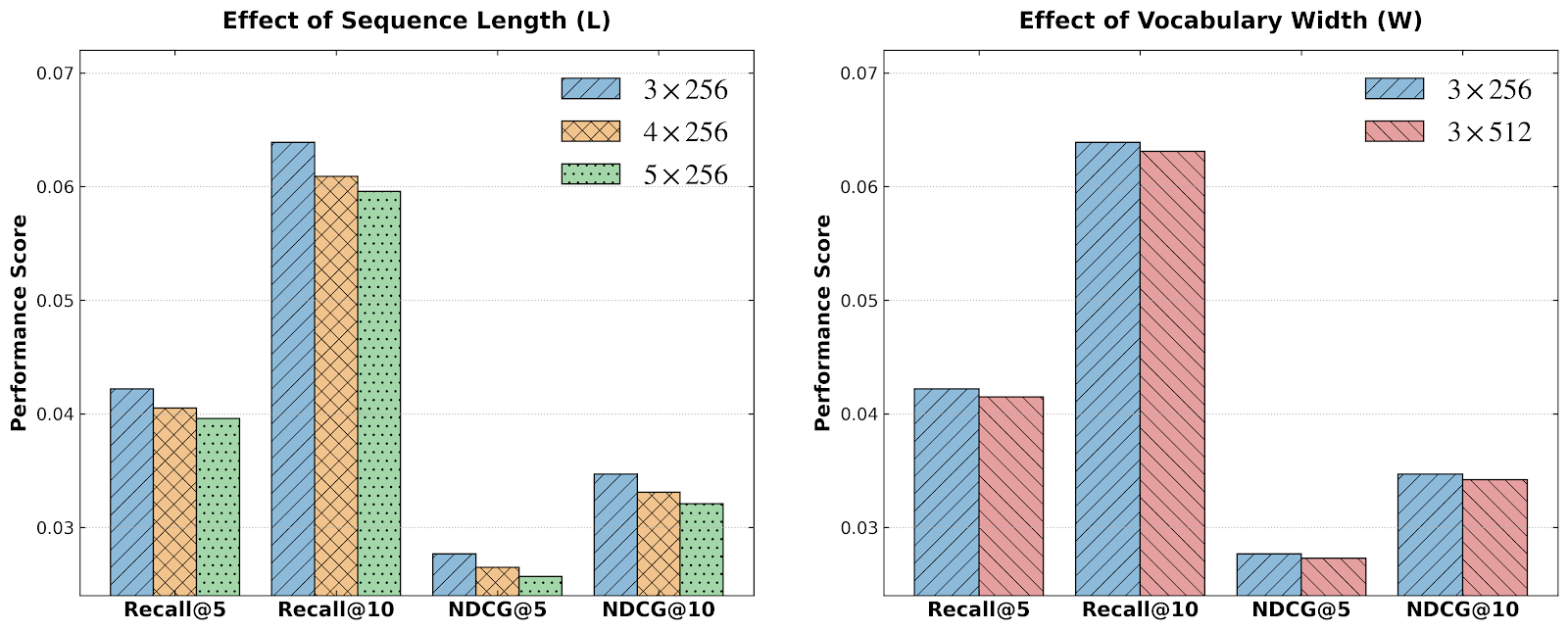}
    \vspace{-0.2cm}
    \caption{The adoption of finer-grained SID representations, entailing longer sequences and larger vocabularies, results in performance degradation.}
    \Description{Two grouped bar charts place Recall@5, Recall@10, NDCG@5, and NDCG@10 on the horizontal axes and performance score on the vertical axes. Left: length-by-width settings 3 by 256, 4 by 256, and 5 by 256 show progressively lower scores as SID length increases. Right: 3 by 512 is slightly below 3 by 256 on all four metrics. The plotted declines are larger for increasing length than for doubling vocabulary width.}
    \label{fig:sid_len_vocab}
\end{figure}

Although SID-GR demonstrates a higher potential compared to item ID-based GR, it also exposes two critical limitations that hinder its practical deployment:
(1) High resource consumption and slow training speed~\cite{forge}. Expanding an item ID into multiple SIDs significantly increases the sequence length, leading to quadratic growth in memory (VRAM) consumption and training time.
(2) Large performance fluctuations and low stability~\cite{unstable,unstable_2,grid,unstable_3}. The relationship between the granularity of SIDs and recommendation accuracy is non-monotonic and exhibits inconsistent patterns across different models and datasets.
Currently, research on SID-GR remains in its early stages. Efforts have prioritized improving SID informativeness~\cite{letter,wang2024eager,xiao2025unger,wang2025generative,wei2025oneloc,forge} and generation speed~\cite{onerec,opq}, while optimizations of the "GR" backbone aimed at mitigating these specific inefficiencies remain relatively under-explored.

We conjecture that these two issues (training inefficiency and performance instability) share a common structural source: both can be traced to the same $L$-fold sequence expansion inherent to SID tokenization, which we term the Semantic Dilution Effect. User intent is typically driven by a sparse subset of critical tokens within SIDs, meaning much of the sequence length consists of non-essential redundancy~\cite{earn,ding2023prune,alvar2025divprune,lin2024not}. This redundancy not only incurs massive computation (Information Non-uniformity)~\cite{sorscher2022beyond,liu2024lost} but also dilutes the already sparse learning signal in recommendation, making it difficult for the model to distinguish pivotal features from noise (Supervision Sparsity)~\cite{bachmann2024pitfalls,bengio2015scheduled}. Unlike general language modeling, the sparse feedback loop in recommendation may be insufficient to regularize such high-granularity, high-redundancy sequences.

Viewed through the lens of Semantic Dilution, existing optimization strategies from adjacent domains reveal inherent limitations. (1) \textit{KV caching} and \textit{Efficient attention kernels}~\cite{attention_1,attention_2,attention_3,H2O,VATP} mitigate the computational burden of long-sequence decoding; however, these are primarily optimized for the long-horizon inference of complex reasoning tasks, whereas SID-GR is severely constrained by the massive retraining overhead required to capture volatile user interests. (2) \textit{Token pruning}~\cite{ye2025fit,zhan2024exploring,ding2023prune,alvar2025divprune,LTP,TM,Dynamicvit} aims to skip redundant inputs, but unlike NLP where non-informative particles (e.g., stop words) are distinct, every token in a high-granularity SID is a semantically dense identifier with complex inter-dependencies, making it difficult to distinguish "salient" tokens from noise without compromising representation quality~\cite{earn}. Consequently, effective pruning mechanisms tailored for SID-GR remain largely unexplored.

We propose \textbf{STAMP} (\textbf{S}emantic \textbf{T}rimming and \textbf{A}uxiliary \textbf{M}ulti-step \textbf{P}rediction), a dual-end framework addressing low input information density and sparse output supervision to counteract the Semantic Dilution Effect.
On the input side, Semantic Adaptive Pruning (SAP) addresses Information Non-uniformity by assessing token utility via semantic features and attention centrality, removing redundancy during the forward pass~\cite{H2O,VATP}. This reduces computational overhead through compact, information-rich representations. SAP builds on our prior work RASTP~\cite{zhan2026rastp}, which preliminarily explored this pruning strategy on the T5 backbone only; here we generalize it to decoder-only architectures, couple it with output-side supervision, and evaluate it on a large-scale industrial dataset.
On the output side, we incorporate Multi-step Auxiliary Prediction (MAP) to address Supervision Sparsity. Beyond standard next-token prediction, MAP employs a multi-token objective~\cite{better} that densifies the feedback loop. It strengthens long-range dependency learning and provides robust signals with compressed inputs~\cite{better,deepseek_v3}. By coupling input purification with signal amplification, STAMP jointly targets training efficiency and representation robustness. Our ablation study examines whether the modules act independently or reinforce one another.
We implemented STAMP on two representative architectures: Encoder-Decoder (T5) and Decoder-Only (Qwen). Experiments on Amazon datasets and a large-scale industrial dataset (AL-GR-Tiny) demonstrate its effectiveness. Results show that STAMP achieves a 1.23--1.38$\times$ speedup and reduces VRAM usage by 17.2\%--54.7\% while maintaining recommendation performance.

The main contributions of this work are summarized as follows,
\begin{itemize}[leftmargin=1.6em, labelsep=0.6em, itemsep=0.2em, topsep=2pt]
\item We propose the Semantic Dilution Effect as a unifying perspective on the inefficiency and instability of SID-GR, supported by indirect empirical evidence from pruning tolerance and attention analysis, guiding dynamic SID learning.
\item We propose STAMP, a dual-end framework that couples input purification with signal amplification to reconcile training efficiency with model robustness.
\item Comprehensive experiments across multiple architectures and large-scale datasets validate STAMP as a viable solution for industrial generative recommendation.
\end{itemize}

\section{Related Work}

\noindent \textbf{Generative Retrieval.}
Generative recommendation has recently emerged~\cite{tiger,gr_1,gr_2,gr_3,tsgr,yu2026thinkrec}. Prior to this shift, recommendation was dominated by discriminative retrieval models~\cite{wang2026pi2i}, with LLMs primarily used as auxiliary components to enhance them~\cite{chen2025enhancing,tan2024idgenrec,ren2024enhancing}. In contrast, the generative retrieval framework serializes item interaction sequences into discrete tokens, allowing the model to directly predict target items in an autoregressive manner~\cite{forge,earn,grid}. Subsequent studies have focused on refining the learning process by developing optimized training frameworks~\cite{rw_14,rw_15_22} to handle large-scale data, while strategies such as reinforcement learning and preference optimization~\cite{zheng2025egav2,wang2024reinforcement,fu2026rankgr} have been integrated to better align generated outputs with complex user preferences.

\noindent \textbf{Semantic Identifier.}
Generative recommendation depends on tokenization. The SID paradigm, pioneered by TIGER~\cite{tiger}, represents items as hierarchical codeword sequences that enable knowledge transfer among semantically similar entities. While early methods derived codes purely from content modalities, subsequent work~\cite{rw_21,rw_15_22,rw_24,ling2026beyond} argues content alone is insufficient and incorporates collaborative signals. Regarding encoding efficiency~\cite{lin2025order}, OneRec~\cite{onerec} streamlines quantization via Residual Quantized K-means, whereas RPG~\cite{rpg} replaces residual quantization with Optimized Product Quantization~\cite{opq} to enable parallel SID prediction. However, it is structurally incompatible with the RQ-based SID construction used in many scenarios. GRID~\cite{grid} introduced a unified, open-source SID-GR framework, while FORGE~\cite{forge} targeted industrial deployment with a large-scale benchmark.

\noindent \textbf{Acceleration of LLMRec.}
Several approaches address generative recommendation efficiency. Regarding model complexity, techniques such as quantization~\cite{gptq,awq,liu2025chord}, knowledge distillation~\cite{distillation_1,distillation_2}, and customized model slimming or adaptation~\cite{fu2024diet,fu2025forward,liu2026race} effectively reduce memory and inference cost, but they mainly target deployment-time efficiency and usually require additional post-training or customization phases. To target latency reduction, researchers have explored speculative sampling~\cite{decode_1, decode_2} and efficient attention mechanisms~\cite{attention_1, attention_2, attention_3, fuxi}. More recently, token pruning has been investigated to skip redundant computation in NLP and CV~\cite{earn, ye2025fit, zhan2024exploring, ding2023prune, alvar2025divprune,beacon,zhan2026rastp}. Despite their potential, speculative sampling and existing token pruning methods remain primarily inference-centric, leaving training-time efficiency comparatively under-explored.

\section{Preliminaries}
\label{sec:preliminaries}

\subsection{SID-based Generative Recommendation}

\begin{figure}
    \centering
    \includegraphics[width=0.98\linewidth]{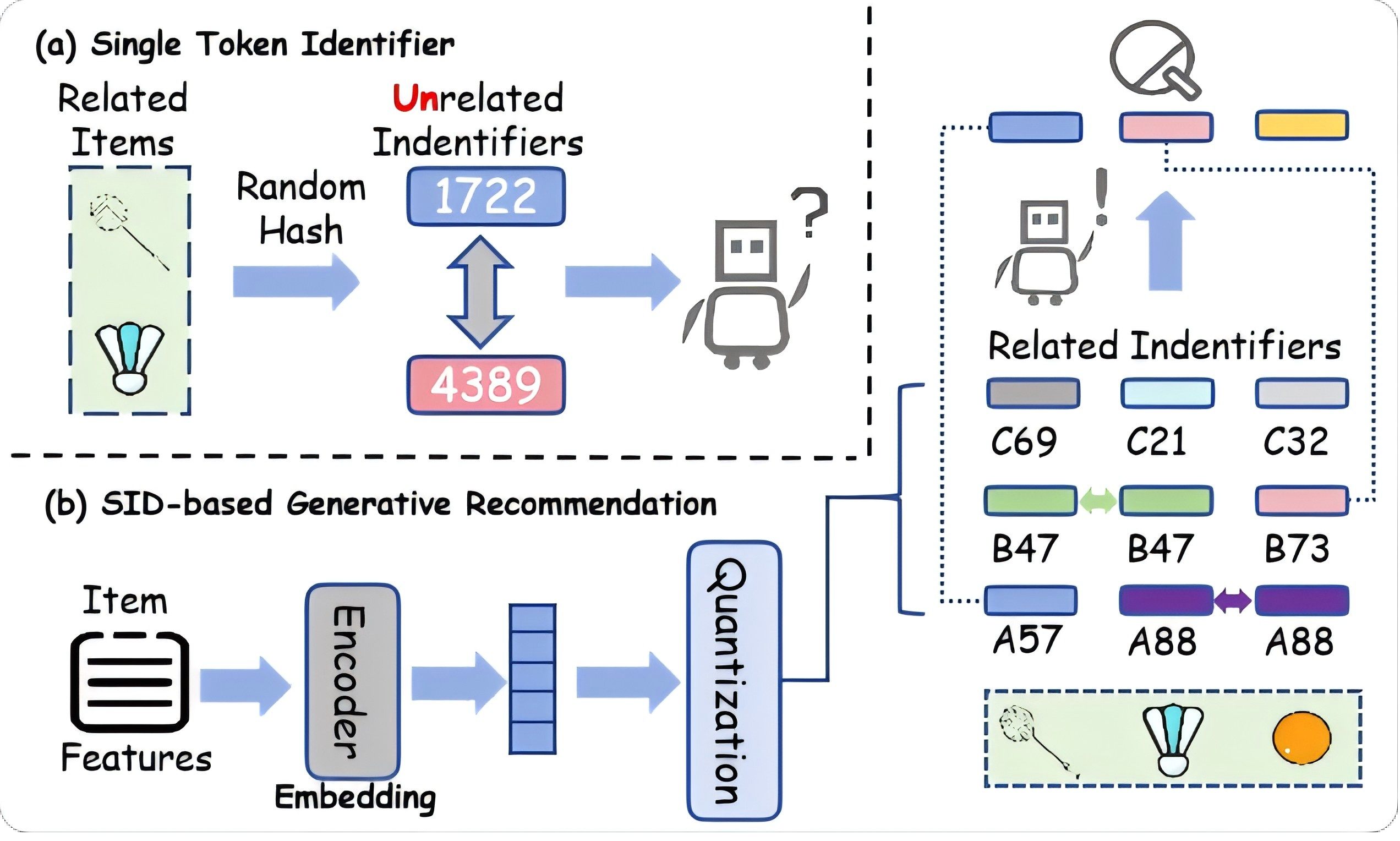}
    \vspace{-0.2cm}
    \caption{Comparison between (a) the single-token identifier mechanism and (b) the SID-based generative recommendation framework.}
    \Description{Panel (a) maps a racket and shuttlecock through random hashing to unrelated atomic IDs 1722 and 4389. Panel (b) sends item features through an encoder, an embedding, and quantization to three-token codes for a racket, shuttlecock, and ball. The racket and shuttlecock share B47, while the shuttlecock and ball share A88; their remaining codes differ. Dotted links connect selected codes to a generated three-token identifier.}
    \label{fig:sid_gr}
\end{figure}

\noindent\textbf{Generative Recommendation.}~
GR reformulates recommendation as an end-to-end sequence generation problem rather than a ranking task. Given a user $u \in \mathcal{U}$ with chronological history $H_u = [i_1, \dots, i_t]$, GR learns a model $P_\theta$ that directly generates the identifier of the target item $i_{t+1}$ autoregressively, aligning recommendation with language modeling and its associated LLM capabilities.

\noindent\textbf{Semantic ID Tokenization.}~
Independent atomic IDs suffer from vocabulary explosion and the cold-start problem. SIDs address this via a two-stage encoding-then-quantization process:
\begin{equation}
s_i = \mathcal{T}(i) = \text{RQ}(h_i) = [c_{i,1}, c_{i,2}, \dots, c_{i,L}],
\end{equation}
where $h_i$ is the semantic embedding of item $i$ (e.g., from Flan-T5-XL~\cite{flan-t5}) and $\text{RQ}(\cdot)$ is a hierarchical residual quantizer (RQ-VAE~\cite{rqvae} or RQ-Kmeans~\cite{onerec}) mapping it to a length-$L$ code over a shared, compact codebook $\mathcal{V}$. This indexes massive item spaces without vocabulary expansion while letting semantically similar items share tokens, mitigating cold-start~\cite{lin2025order}.

\noindent\textbf{SID-based Generative Recommendation.}~
The model serializes the user's history into $X = [\mathcal{T}(i_1), \dots, \mathcal{T}(i_t)]$, an input of length $N = t \times L$, and is trained to autoregressively predict the target's SID sequence $Y=[y_1,\dots,y_L]$ by minimizing:
\begin{equation}
\mathcal{L}_{\text{NTP}} = - \sum_{j=1}^{L} \log P(y_j \mid X, y_{<j}; \theta).
\label{eq:2}
\end{equation}

We posit that both challenges are related to the same $L$-fold expansion: decomposing an item into $L$ SIDs simultaneously (i) inflates $X$ with tokens of low marginal information density, and (ii) requires $L$ consecutive predictions under a per-example supervisory budget that does not scale with $L$. Sequence length thus outgrows input information content and supervision density.

\begin{figure}
    \centering
    \includegraphics[width=0.98\linewidth]{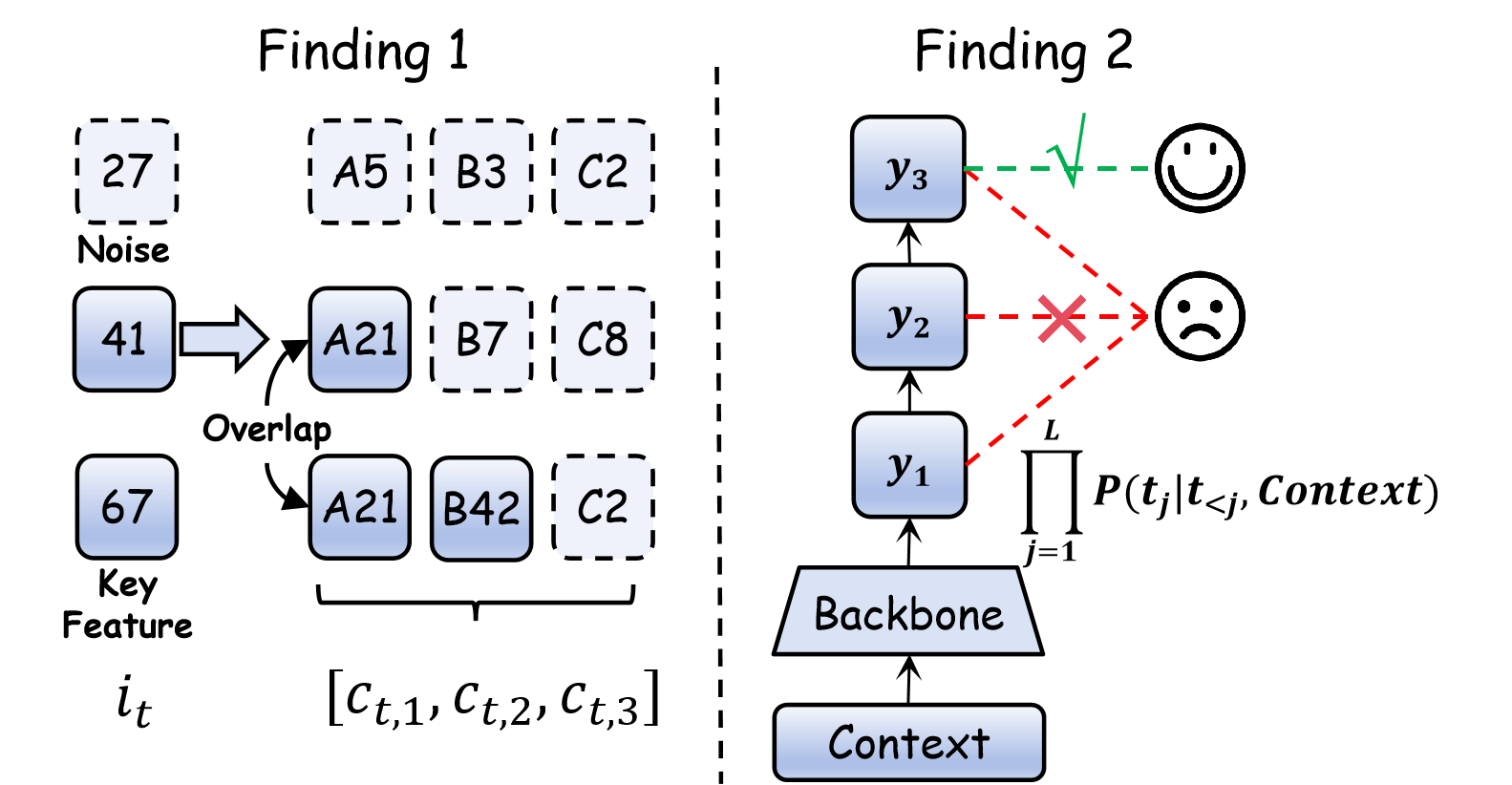}
    \caption{Redundancy is characterized by feature-level redundancy and sequence-level semantic overlaps (left). The joint probability constraint imposes higher demands on the model's long-range dependency capabilities (right).}
    \Description{Left: item IDs 27, 41, and 67 are expanded into three-token SID rows. Dashed boxes mark illustrative noise or redundant features, solid boxes mark selected key features, and an overlap arrow links the repeated A21 token in the last two rows. Right: context feeds a backbone that generates y1, y2, and y3 sequentially. A product of conditional token probabilities accompanies a green success path and red failure paths, illustrating how an incorrect token can prevent the full identifier from matching the target. These are conceptual examples, not measured results.}
    \label{fig:findings}
\end{figure}

\subsection{Finding 1: Information Non-uniformity}
Non-critical, redundant information pervades machine learning: the Lottery Ticket Hypothesis~\cite{frankle2018lottery} demonstrates that most parameters in dense networks are functionally redundant; token pruning and KV-cache eviction methods~\cite{ding2023prune,alvar2025divprune,earn,ye2025fit,xiao2024efficient,zhan2024exploring} discard many tokens without damaging performance; near-duplicate training samples yield rapidly diminishing returns~\cite{sorscher2022beyond}; and LLMs struggle to uniformly exploit information dispersed across long, redundant contexts~\cite{liu2024lost}. The common thread is that model input information density is highly non-uniform: a small subset of elements carries the bulk of the predictive signal, while the rest is redundant or distracting.

This effect is structurally amplified in SID-GR. Decomposing one item into $L$ tokens boosts zero-shot semantic expressiveness but proportionally expands the input with tokens of low marginal information density, manifesting as two interconnected redundancy types: (1) \textit{Feature-level Redundancy}, where user intent is often driven by a few key features (e.g., $\text{Brand}$) rather than the full hierarchy $\text{Category}\rightarrow\text{Brand}\rightarrow\text{Attribute}$, so fine-grained attributes act as comparatively irrelevant noise; (2) \textit{Sequence-level Redundancy}, where historical items frequently share identical SIDs (e.g., brand, category), so repeated encoding yields diminishing marginal information across the temporal dimension.

Repeated tokens can still be informative: a recurring brand token is a valid preference signal. The \textit{first} occurrence may carry greater information gain; subsequent repetitions may mainly reinforce an already encoded signal with lower marginal information gain. Indiscriminately retaining repeated SIDs spreads the limited attention budget across near-duplicates rather than distinctive tokens that drive prediction.

Notably, this redundancy also carries an optimization opportunity. SID decouples an item's entangled attributes into independent, discrete tokens, allowing attention to evaluate each token's contribution in isolation rather than disentangle redundancy from a monolithic item embedding. This structural separation gives pruning a natural handle: low-utility tokens can be located and removed at the granularity of individual SIDs, enabling aggressive compression while preserving the specific tokens that carry the user's core intent.

\subsection{Finding 2: Supervision Sparsity}
Unlike general generative tasks where every generated token is an immediate supervision signal, GR derives valid training signal solely from the ground-truth item at the sequence's end, forcing gradients to backpropagate across long contexts without intermediate guidance. Although SID extends the prediction target from a single atomic ID to $L$ tokens, the input sequence expands by the same factor $L$, so the \textit{relative} density of supervision per input token does not improve: feedback still comes from one training example's worth of ground truth, spread across a proportionally longer context. This makes it substantially harder for the model to attribute the final purchase decision to the specific, sparse set of key features embedded within an otherwise noisy interaction history.

Moreover, correctly predicting an item requires all $L$ SID tokens to be right consecutively: each token's prediction conditions on all preceding tokens, so a single early mistake can shift the remaining trajectory away from the intended item:
\begin{equation}
    P(\text{Item}) = \prod_{j=1}^{L} P(y_j \mid y_{<j}, \text{Context})
\end{equation}
so early-token errors compound via the chain rule, causing degradation as $L$ grows~\cite{grid,forge,bachmann2024pitfalls}. Unlike open-ended language generation, where numerous plausible continuations can absorb this effect, SID-GR admits only one exact target sequence per interaction, leaving the sparse, single-point terminal supervision especially fragile and motivating auxiliary supervision to stabilize training and strengthen long-range dependency induction.

\begin{figure*}
    \centering
    \includegraphics[width=0.98\linewidth]{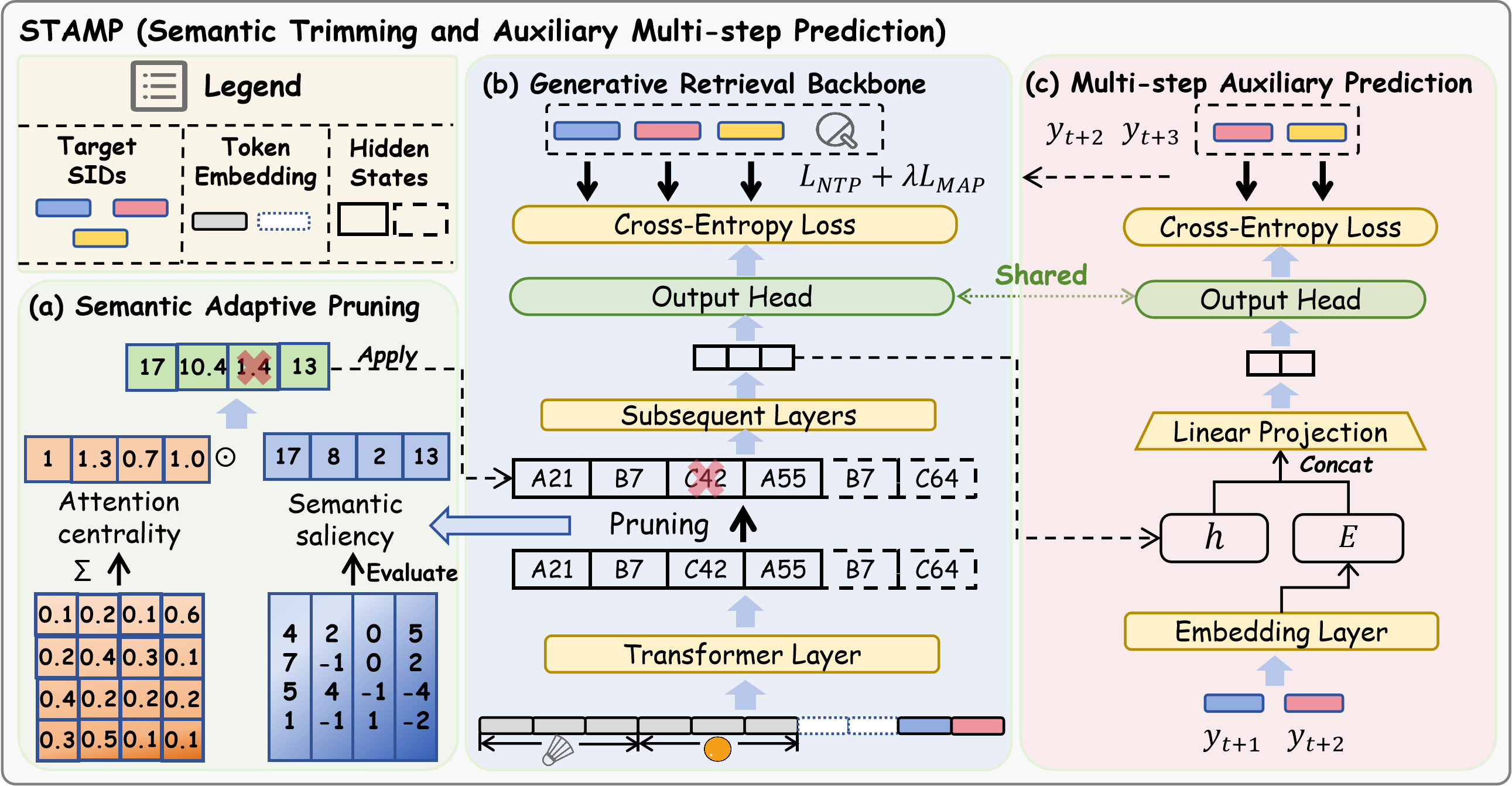}
    \vspace{-0.2cm}
    \caption{Overview of the STAMP framework. (b) STAMP accelerates training by compressing sequences via pruning while ensuring performance through multi-token prediction. (a) SAP computes token importance to retain high-utility information and remove redundancy. (c) MAP predicts multiple future items simultaneously to provide richer supervision signals.}
    \Description{Three connected panels. (a) Attention-column sums and hidden-state magnitudes are multiplied to score tokens; the lowest-scoring token, C42, is crossed out. (b) Item-token embeddings pass through a Transformer layer, pruning, subsequent layers, and an output head with next-token cross-entropy loss. (c) Backbone hidden states are concatenated with embeddings of the following ground-truth tokens, projected, and passed through the shared output head to predict one token further ahead. The primary and auxiliary losses are combined with weight lambda.}
\label{fig:framework}
\end{figure*}

\section{Methodology}
\subsection{Overview}
\textbf{STAMP} (Semantic Trimming and Auxiliary Multi-step Prediction) integrates two modules into the standard Transformer backbone to balance training efficiency and representation robustness (Figure~\ref{fig:framework}). \textbf{SAP}, inserted at layer $l_{prune}$, evaluates token importance through intrinsic information and extrinsic structural centrality, removing redundant tokens to accelerate subsequent layers. \textbf{MAP} densifies supervision by predicting future tokens, enhancing long-range dependency capture. Both modules are lightweight: SAP reuses the backbone's attention weights, and MAP uses a single MLP with a shared LM head, introducing few additional parameters.

\subsection{Semantic Adaptive Pruning}
We design SAP by capitalizing on the Transformer's intrinsic self-attention mechanism. Since attention weights naturally quantify the information flow between tokens, they serve as an effective, computationally efficient proxy for token utility~\cite{H2O,VATP}, enabling SAP to dynamically identify and retain discriminative SIDs without incurring significant additional computational overhead. We reuse the hidden states and attention weights already produced at the pruning layer, rather than adding another backbone pass. The resulting scores can therefore guide token retention using information available within the existing computation.

\subsubsection{\textbf{Token Importance Evaluation}} Let $\mathbf{H}^{(l)} \in \mathbb{R}^{B \times N \times d}$ be the hidden states at pruning layer $l$. We score each token along two complementary dimensions. \textbf{Semantic Saliency} uses the $\ell_1$-norm, since high-magnitude features typically indicate significant semantic activation~\cite{Dynamicvit}:
\begin{equation}S_{\text{sem}}(i) = \text{Norm}\left(|\mathbf{h}^{(l)}_i|_1\right),\end{equation}
where $\text{Norm}(\cdot)$ stabilizes the scores. \textbf{Attention Centrality}~\cite{shang2025llava} captures token $i$'s structural importance by summing the attention it receives from all other tokens and heads:
\begin{equation}S_{\text{attn}}(i) = \sum_{h=1}^{\mathcal{H}} \sum_{j=1}^{N} \mathbf{A}_{h, j, i},\end{equation}
where a higher score marks $i$ as a frequently-attended information hub. The final importance score multiplies both signals:
\begin{equation}\mathcal{I}_i = S_{\text{sem}}(i) \times S_{\text{attn}}(i),\end{equation}
so a high $\mathcal{I}_i$ requires a token to be both internally rich and externally pivotal.

\subsubsection{\textbf{Architecture-Adaptive Pruning}}
We shorten subsequent layers' input using order-preserving sequence compression. Importance scores determine which tokens survive, not their order. This preserves the temporal structure of user interactions.

\textbf{Token Selection.} The execution begins by identifying the set of indices, $\mathcal{K}$, intended for retention. We retain the top-scoring indices with $|\mathcal{K}| = \lfloor \alpha \cdot N \rfloor$ under a retention ratio $\alpha \in (0,1]$ that determines the total budget of preserved tokens. For decoder-only architectures (e.g., Qwen), where input history and the generation target share one token stream, we additionally protect a window $\mathcal{W}_{\text{recent}} \subseteq \mathcal{K}$ spanning the target segment and its immediate boundary, retained regardless of importance score to preserve gradient flow to the target; the remaining budget in $\mathcal{K}$ is filled by the highest-scoring tokens elsewhere.

\textbf{Index Re-ordering and Compression.} Naively gathering selected tokens by importance would disrupt the temporal integrity of the user interaction sequence. Therefore, we first sort the selected indices to restore their original relative positions:
\begin{equation}
    \mathcal{K}_{\text{sorted}} = \operatorname{Sort}(\mathcal{K}, \text{order}=\text{ascending}).
\end{equation}
Subsequently, we apply compression by gathering the hidden states corresponding to these sorted indices:
\begin{equation}
    H_{\text{compressed}}^{(l)} = \operatorname{Gather}\left(H^{(l)}, \mathcal{K}_{\text{sorted}}\right).
\end{equation}
This operation reduces the sequence length from $N$ to $\alpha \cdot N$ for all subsequent layers. Surviving tokens retain their \textit{original} absolute position identifiers rather than being renumbered, so relative or rotary positional encodings (e.g., Qwen) still perceive the true temporal spacing between interactions rather than misreading pruned-adjacent tokens as consecutive.

\subsection{Multi-step Auxiliary Prediction}
MAP addresses supervision dilution via Multi-Token Prediction (MTP)~\cite{better}. Given SIDs are typically only 3 tokens, we extend the supervision horizon with a single auxiliary head.

\subsubsection{\textbf{Dense Supervision Formulation}}
We add an objective forecasting $y_{t+2}$ alongside the standard target $y_{t+1}$. Given backbone state $\mathbf{h}_t \in \mathbb{R}^d$, we fuse it with the ground-truth embedding $E(y_{t+1})$ into a \textit{foresight representation}:
\begin{equation}
    \mathbf{h}^{mtp}_t = \operatorname{MLP}\left(\operatorname{Concat}(\mathbf{h}_t, E(y_{t+1}))\right),
\end{equation}
which simulates the step-$(t+1)$ hidden state and forces the backbone to encode features transferable to future contexts. Conditioning on the ground-truth $y_{t+1}$, rather than a possibly-erroneous self-prediction, mirrors how DeepSeek-V3's MTP module keeps its prediction chain causally consistent across depths~\cite{deepseek_v3}, unlike the fully independent, parallel heads of earlier MTP formulations~\cite{better}; this keeps the auxiliary objective cheap (one MLP, shared LM head) while still probing $\mathbf{h}_t$'s ability to encode generalizable, forward-looking features.

\subsubsection{\textbf{Shared Projection Strategy}}
We map $\mathbf{h}^{mtp}_t$ to the vocabulary space using the backbone's own unembedding matrix rather than a new head:
\begin{equation}
    P(y_{t+2} \mid \mathbf{h}_t, y_{t+1}) = \operatorname{Softmax}(\operatorname{LM\_Head}(\mathbf{h}^{mtp}_t)),
\end{equation}
aligning the auxiliary representation with the primary semantic space at no extra parameter cost.

\subsection{Optimization Objective}
STAMP is trained end-to-end with the primary NTP loss (Eq.~\ref{eq:2}) plus the MAP objective, which minimizes the negative log-likelihood of $y_{j+1}$ given the foresight context at step $j$ (summed to $L-1$, since predicting $y_{j+1}$ requires ground-truth $y_j$ as input):
\begin{equation}
    \mathcal{L}_{\text{MAP}} = - \sum_{j=1}^{L-1} \log P(y_{j+1} \mid \mathbf{h}_j, y_j).
\end{equation}
The overall training objective is then a weighted sum of the primary and auxiliary losses:
\begin{equation}
    \mathcal{L}_{\text{total}} = \mathcal{L}_{\text{NTP}} + \lambda \cdot \mathcal{L}_{\text{MAP}},
\end{equation}
where $\lambda$ balances the auxiliary signal. As a training-only objective, MAP is discarded at inference, incurring zero latency overhead.

\section{Experiments}

\begin{sloppypar}
We evaluate STAMP across architectures and data scales using two frameworks: \textbf{GRID}~\cite{grid} implements the Encoder-Decoder (T5) architecture on three public Amazon datasets, while \textbf{FORGE}~\cite{forge} instantiates the Decoder-Only (Qwen) architecture on the industrial dataset AL-GR-Tiny. This setup addresses the following research questions:
\end{sloppypar}

\begin{itemize}[leftmargin=1.6em, labelsep=0.6em, itemsep=0.2em, topsep=2pt]
\item \textbf{RQ1:} How does STAMP perform in performance and efficiency on different architectures of SID-based generative recommendation?
\item \textbf{RQ2:} What roles do STAMP’s two modules, SAP and MAP, play in its performance?
\item \textbf{RQ3:} How do the design configurations of SAP (\textit{i.e.}, pruning strategy, pruning layer, and retention ratio) affect the model performance?
\item \textbf{RQ4:} What mechanism enables STAMP to safely prune the majority of tokens?
\end{itemize}

\subsection{Experimental Settings}
\subsubsection{\textbf{Datasets}} We employ four datasets divided into two categories, summarized in Table \ref{tab:dataset_stats}.
\begin{table}[h]
    \centering
    \caption{Statistics of the datasets.}
    \label{tab:dataset_stats}
    \vspace{-0.2cm}
    \setlength{\tabcolsep}{16pt}
    \resizebox{\linewidth}{!}{
    \begin{tabular}{lrrr}
        \toprule[1.5pt]
        \textbf{Dataset} & \textbf{\#Users} & \textbf{\#Items} & \textbf{\#Interactions} \\
        \midrule
        Sports & 35.6K & 18.4K & 296.3K \\
        Beauty & 22.4K & 12.1K & 198.5K \\
        Toys  & 19.4K & 11.9K & 167.6K \\
        AL-GR & 131M & 251M & 14B \\
        \bottomrule[1.5pt]
    \end{tabular}
    }
\end{table}
\begin{table*}[h]
    \caption{Overall performance on GRID (T5) over Beauty, Sports and Toys. Results are averaged over five random seeds to mitigate performance variance caused by streaming training and early stopping. We report the training time and VRAM usage on a single NVIDIA A100 GPU. $\uparrow$ indicates higher is better, and $\downarrow$ indicates lower is better.}
    \vspace{-0.2cm}
    \centering
    \setlength{\tabcolsep}{6pt}
    \resizebox{1.0\linewidth}{!}{
    \begin{tabular}{c|c|c|cccc|cccc}
    \toprule[1.5pt]
    \multirow{2}{*}{Backbone} & \multirow{2}{*}{Dataset} & \multirow{2}{*}{Method} & \multicolumn{4}{c|}{Performance Metrics} & \multicolumn{4}{c}{Efficiency Metrics} \\
    \cmidrule{4-11}
     & & & Recall@5 & Recall@10 & NDCG@5 & NDCG@10 & Time $\downarrow$ & Speedup $\uparrow$ & VRAM $\downarrow$ & Reduction $\uparrow$ \\
    \midrule
     & & Base & 0.0426 \scriptsize{$\pm$ 0.0013} & 0.0645 \scriptsize{$\pm$ 0.0019} & 0.0282 \scriptsize{$\pm$ 0.0014} & 0.0353 \scriptsize{$\pm$ 0.0013} & 212s & -- & 22234MiB & -- \\
     & & EARN (L=2) & 0.0404 \scriptsize{$\pm$ 0.0013} & 0.0619 \scriptsize{$\pm$ 0.0010} & 0.0263 \scriptsize{$\pm$ 0.0012} & 0.0333 \scriptsize{$\pm$ 0.0010} & 116s & 1.83$\times$ & 13544MiB & 39.1\% \\
    \rowcolor{gray!15} \cellcolor{white} & \cellcolor{white} & STAMP (L=2) & \textbf{0.0444} \scriptsize{$\pm$ 0.0012} & 0.0655 \scriptsize{$\pm$ 0.0015} & \textbf{0.0295} \scriptsize{$\pm$ 0.0011} & \textbf{0.0363} \scriptsize{$\pm$ 0.0012} & 170s & 1.25$\times$ & 13862MiB & 37.7\% \\
     & & EARN (L=1) & 0.0379 \scriptsize{$\pm$ 0.0010} & 0.0591 \scriptsize{$\pm$ 0.0009} & 0.0246 \scriptsize{$\pm$ 0.0006} & 0.0315 \scriptsize{$\pm$ 0.0005} & \textbf{98s} & \textbf{2.16$\times$} & \textbf{9116MiB} & \textbf{59.0\%} \\
    \rowcolor{gray!15} \cellcolor{white} & \cellcolor{white}\multirow{-5}{*}{Beauty} & STAMP (L=1) & 0.0441 \scriptsize{$\pm$ 0.0010} & \textbf{0.0657} \scriptsize{$\pm$ 0.0022} & 0.0293 \scriptsize{$\pm$ 0.0006} & \textbf{0.0363} \scriptsize{$\pm$ 0.0009} & 154s & 1.38$\times$ & 10378MiB & 53.3\% \\
    \cmidrule{2-11}
     & & Base & \textbf{0.0234} \scriptsize{$\pm$ 0.0004} & \textbf{0.0354} \scriptsize{$\pm$ 0.0004} & \textbf{0.0154} \scriptsize{$\pm$ 0.0002} & \textbf{0.0193} \scriptsize{$\pm$ 0.0001} & 205s & -- & 29370MiB & -- \\
     & & EARN (L=2) & 0.0208 \scriptsize{$\pm$ 0.0015} & 0.0322 \scriptsize{$\pm$ 0.0018} & 0.0135 \scriptsize{$\pm$ 0.0012} & 0.0172 \scriptsize{$\pm$ 0.0013} & 114s & 1.80$\times$ & 17834MiB & 39.3\% \\
    \rowcolor{gray!15} \cellcolor{white} & \cellcolor{white} & STAMP (L=2) & 0.0228 \scriptsize{$\pm$ 0.0015} & 0.0352 \scriptsize{$\pm$ 0.0015} & 0.0153 \scriptsize{$\pm$ 0.0008} & \textbf{0.0193} \scriptsize{$\pm$ 0.0008} & 166s & 1.23$\times$ & 18318MiB & 37.6\% \\
     & & EARN (L=1) & 0.0197 \scriptsize{$\pm$ 0.0009} & 0.0311 \scriptsize{$\pm$ 0.0014} & 0.0126 \scriptsize{$\pm$ 0.0006} & 0.0163 \scriptsize{$\pm$ 0.0007} & \textbf{95s} & \textbf{2.16$\times$} & \textbf{12074MiB} & \textbf{58.9\%} \\
    \rowcolor{gray!15} \cellcolor{white} & \cellcolor{white}\multirow{-5}{*}{Sports} & STAMP (L=1) & 0.0232 \scriptsize{$\pm$ 0.0009} & \textbf{0.0354} \scriptsize{$\pm$ 0.0019} & 0.0153 \scriptsize{$\pm$ 0.0009} & 0.0192 \scriptsize{$\pm$ 0.0012} & 151s & 1.36$\times$ & 13302MiB & 54.7\% \\
    \cmidrule{2-11}
     & & Base & 0.0345 \scriptsize{$\pm$ 0.0010} & 0.0499 \scriptsize{$\pm$ 0.0008} & 0.0238 \scriptsize{$\pm$ 0.0009} & 0.0287 \scriptsize{$\pm$ 0.0006} & 216s & -- & 22268MiB & -- \\
     & & EARN (L=2) & 0.0312 \scriptsize{$\pm$ 0.0010} & 0.0471 \scriptsize{$\pm$ 0.0017} & 0.0205 \scriptsize{$\pm$ 0.0010} & 0.0256 \scriptsize{$\pm$ 0.0011} & 117s & 1.85$\times$ & 11772MiB & 47.1\% \\
    \rowcolor{gray!15} \cellcolor{white} & \cellcolor{white} & STAMP (L=2) & \textbf{0.0356} \scriptsize{$\pm$ 0.0009} & 0.0503 \scriptsize{$\pm$ 0.0010} & \textbf{0.0246} \scriptsize{$\pm$ 0.0006} & \textbf{0.0293} \scriptsize{$\pm$ 0.0007} & 174s & 1.24$\times$ & 14256MiB & 36.0\% \\
     & & EARN (L=1) & 0.0274 \scriptsize{$\pm$ 0.0007} & 0.0415 \scriptsize{$\pm$ 0.0013} & 0.0179 \scriptsize{$\pm$ 0.0007} & 0.0225 \scriptsize{$\pm$ 0.0009} & \textbf{98s} & \textbf{2.20$\times$} & \textbf{8428MiB} & \textbf{62.2\%} \\
    \rowcolor{gray!15} \cellcolor{white}\multirow{-15}{*}{GRID (T5)} & \cellcolor{white}\multirow{-5}{*}{Toys} & STAMP (L=1) & 0.0344 \scriptsize{$\pm$ 0.0012} & \textbf{0.0509} \scriptsize{$\pm$ 0.0015} & 0.0235 \scriptsize{$\pm$ 0.0009} & 0.0286 \scriptsize{$\pm$ 0.0013} & 159s & 1.36$\times$ & 10658MiB & 52.1\% \\
    \bottomrule[1.5pt]
    \end{tabular}
    }
    \label{tab:performance_efficiency_grid_t5}
\end{table*}

\begin{table*}[h]
    \caption{Overall performance on FORGE (Qwen2.5-Instruct-0.5B) over AL-GR-Tiny. We report the training time and VRAM usage on two NVIDIA A100 GPUs. $\uparrow$ indicates higher is better, and $\downarrow$ indicates lower is better.}
    \vspace{-0.2cm}
    \centering
    \setlength{\tabcolsep}{6pt}
    \resizebox{1.0\linewidth}{!}{
    \begin{tabular}{c|c|c|cccc|cccc}
    \toprule[1.5pt]
    \multirow{2}{*}{Backbone} & \multirow{2}{*}{Dataset} & \multirow{2}{*}{Method} & \multicolumn{4}{c|}{Performance Metrics} & \multicolumn{4}{c}{Efficiency Metrics} \\
    \cmidrule{4-11} 
     & & & Hit@20 & Hit@100 & Hit@500 & Hit@1000 & Time $\downarrow$ & Speedup $\uparrow$ & VRAM $\downarrow$ & Reduction $\uparrow$ \\
    \midrule
     & & Base & \textbf{0.0209} & \textbf{0.0508} & \textbf{0.1155} & 0.1400 & 1139s & -- & 2*64232MiB & -- \\
     & & EARN (L=6) & 0.0152 & 0.0354 & 0.0839 & 0.1021 & \textbf{434s} & \textbf{2.62$\times$} & \textbf{2*29452MiB} & \textbf{54.1\%} \\
    \rowcolor{gray!15} \cellcolor{white}\multirow{-3}{*}{\shortstack{FORGE\\(Qwen)}} & \cellcolor{white}\multirow{-3}{*}{AL-GR-Tiny} & STAMP (L=6) & 0.0207 & 0.0506 & 0.1149 & \textbf{0.1405} & 851s & 1.34$\times$ & 2*53196MiB & 17.2\% \\
    \bottomrule[1.5pt]
    \end{tabular}
    }
    \label{tab:forge_qwen_algr_tiny}
\end{table*}

\begin{itemize}[leftmargin=1.6em, labelsep=0.6em, itemsep=0.2em, topsep=2pt]
\item \textbf{Amazon\footnote{\url{https://jmcauley.ucsd.edu/data/amazon/}}\label{fn:amazon}:} We use the \textit{Beauty}, \textit{Toys}, and \textit{Sports} Amazon review subsets, which cover different product categories. A 5-core filter retains users and items with at least five interactions. Under leave-one-out evaluation, each user's final interaction is reserved for testing, the penultimate for validation, and the remainder for training.

\item \textbf{AL-GR-Tiny\footnote{\url{https://huggingface.co/datasets/AL-GR/AL-GR-Tiny}}\label{fn:algr}:} We evaluate industrial scalability on AL-GR-Tiny, a subset of AL-GR from a large-scale Chinese e-commerce platform. AL-GR is the first industrial-scale benchmark for SID construction in GR, integrating multimodal (text and image) features across over 250 million items. We use its pre-processed open-source Stage 1 (S1) split: four days of training behaviors and a test set from the following day.
\end{itemize}

\subsubsection{\textbf{Baselines.}}
\begin{itemize}[leftmargin=1.6em, labelsep=0.6em, itemsep=0.2em, topsep=2pt]
\item \textbf{EARN\cite{earn}:} An inference acceleration framework using learnable prefix and suffix register tokens to compress user histories. At $1/4$ of the total network depth, it discards original prompt tokens and computes attention exclusively over register tokens in subsequent layers, reducing computation and KV-cache size. For our comparison, we instantiate EARN during training rather than applying it solely at inference.
\end{itemize}

\subsubsection{\textbf{Evaluation Metrics}}
\begin{itemize}[leftmargin=1.6em, labelsep=0.6em, itemsep=0.4em, topsep=2pt]
    \item \textbf{Recommendation Effectiveness:} For GRID, we report \textbf{Recall@K} and \textbf{NDCG@K} ($K \in \{5, 10\}$). For FORGE, we report \textbf{HitRate@K} ($K \in \{20, 100, 500, 1000\}$), the percentage of cases where the ground-truth item appears in the top-K candidates.
    \item \textbf{Training Efficiency:} \textbf{Training Speedup} (time relative to the baseline over 1,600 steps) and \textbf{VRAM Reduction} (percentage decrease in peak GPU memory).
\end{itemize}

\subsubsection{\textbf{Implementation Details.}}\label{sec:implementation_details}
We align our implementation with the standard configurations of the respective frameworks, adjusting only the framework-specific modules introduced by STAMP.

For GRID, we use Flan-T5-XL for semantic encoding and RK-Means tokenization with three codebooks. The backbone is a lightweight T5 Encoder-Decoder (8 layers: 4 encoder, 4 decoder; hidden dim 128; 6 heads; MLP dim 1,024). We set SAP's retention ratio $\alpha=1/3$ and MAP's fusion coefficient $\lambda=0.3$, optimizing with Adam (lr $1\times10^{-3}$, weight decay $1\times10^{-4}$, dropout 0.15) on a single A100-40GB GPU (batch size 32, sequence length 120), with early stopping (patience 10, validated every 1,600 steps). GRID's streaming training uses early stopping, so step counts and final performance vary across methods, runs, and random seeds. We therefore average results over five seeds ($\{1, 42, 999, 1024, 2025\}$).

For FORGE, we use the pre-processed AL-GR-Tiny dataset and a Qwen2.5-0.5B-Instruct backbone with a 1,280-token input (1,024 source, 256 target). We set $\alpha=0.6$ and $\lambda=0.3$, fine-tuning in bfloat16 on two A100-80GB GPUs with AdamW ($\beta_1{=}0.9$, $\beta_2{=}0.999$, $\epsilon{=}1\text{e-}8$, per-device batch size 80) for a single epoch, totaling 100,000 steps. On this large-scale dataset, one epoch suffices for convergence. The fixed-epoch setup keeps training steps identical across compared methods; unlike GRID, we follow FORGE's default single-run configuration without seed-averaging.

\subsection{Overall Performance (RQ1)}
We compare STAMP against EARN across both paradigms. The notation $L=n$ denotes pruning/compression after the $n$-th Transformer layer; following EARN's own layer-ratio convention, we fix $L$ \textit{a priori} at roughly one-quarter and one-half of backbone depth for the aggressive and conservative configurations ($L\in\{1,2\}$ of 4 T5 encoder layers; $L=6$ of 24 Qwen layers).

\subsubsection{\textbf{Performance on GRID (T5)}}
Table \ref{tab:performance_efficiency_grid_t5} compares the base model, baselines, and STAMP.
\begin{itemize}[leftmargin=1.6em, labelsep=0.6em, itemsep=0.2em, topsep=2pt]
    \item \textbf{Efficiency:} SAP's early pruning yields substantial gains: the aggressive configuration ($L=1$) achieves a 1.36--1.38$\times$ speedup and over 52\% VRAM reduction across all datasets. EARN is faster still, but STAMP's acceleration remains highly practical for resource-constrained deployment.
    \item \textbf{Effectiveness:} EARN's aggressive compression (encoding the entire user history into only two register tokens) creates an information bottleneck that causes severe performance degradation. STAMP, in contrast, closely tracks base-model accuracy on all three datasets, with the $L=2$ configuration surpassing it outright on \textit{Beauty} and \textit{Toys}, suggesting SAP acts as a denoising filter that sharpens focus on core preferences. On \textit{Sports}, the base model remains marginally ahead on most metrics, indicating a residual pruning cost. STAMP narrows SAP-only pruning's accuracy gap and incurs a much smaller loss than EARN. MAP's densified supervision further improves training stability and helps offset this residual gap, yielding a favorable speed-accuracy balance overall.
\end{itemize}

\begin{figure}[!t]
    \centering
    \begin{subfigure}[b]{0.495\linewidth}
        \includegraphics[width=\textwidth]{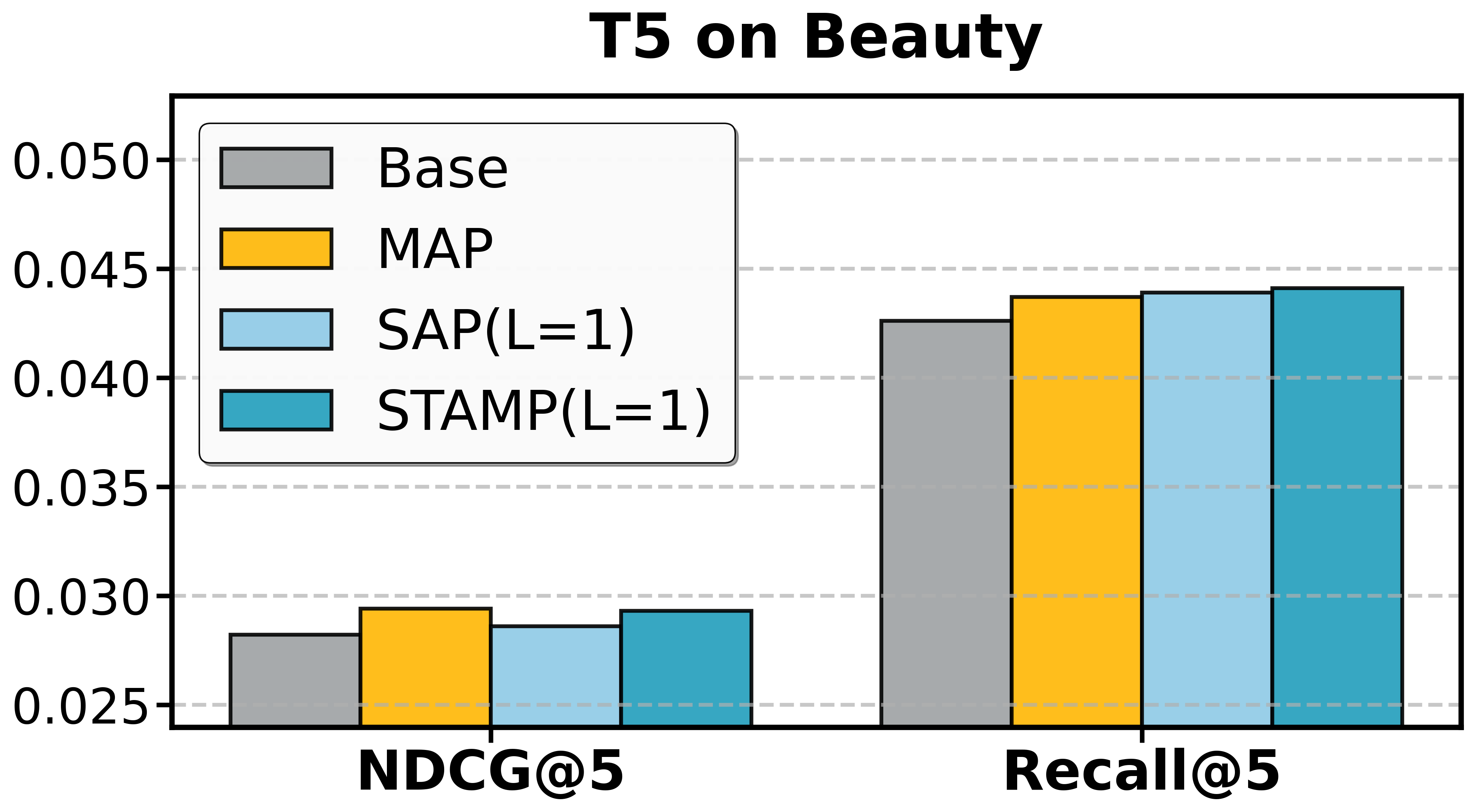}
    \end{subfigure}
    \hfill 
    \begin{subfigure}[b]{0.495\linewidth}
        \includegraphics[width=\textwidth]{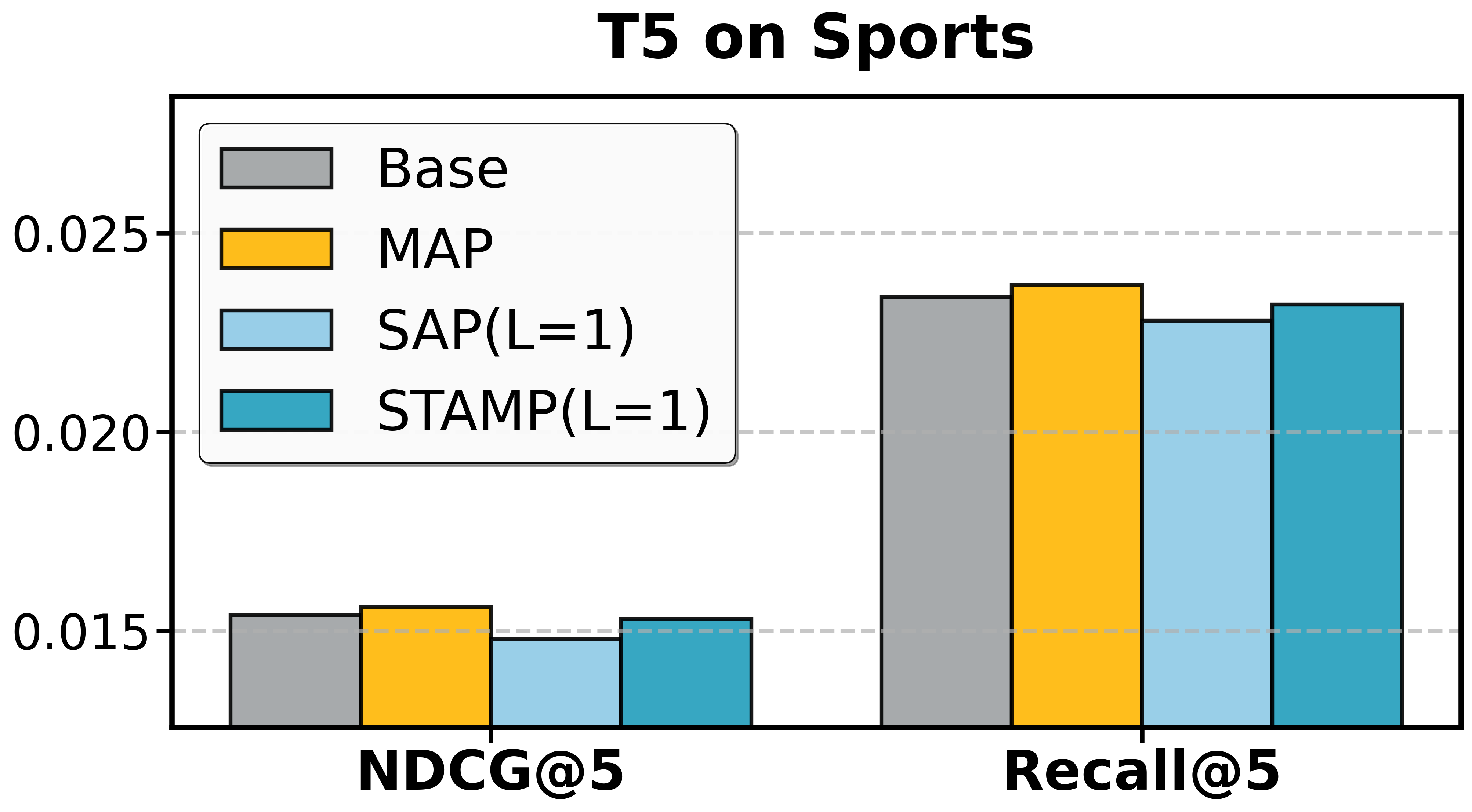}
    \end{subfigure}

    \vspace{0.1cm} 

    \begin{subfigure}[b]{0.495\linewidth}
        \includegraphics[width=\textwidth]{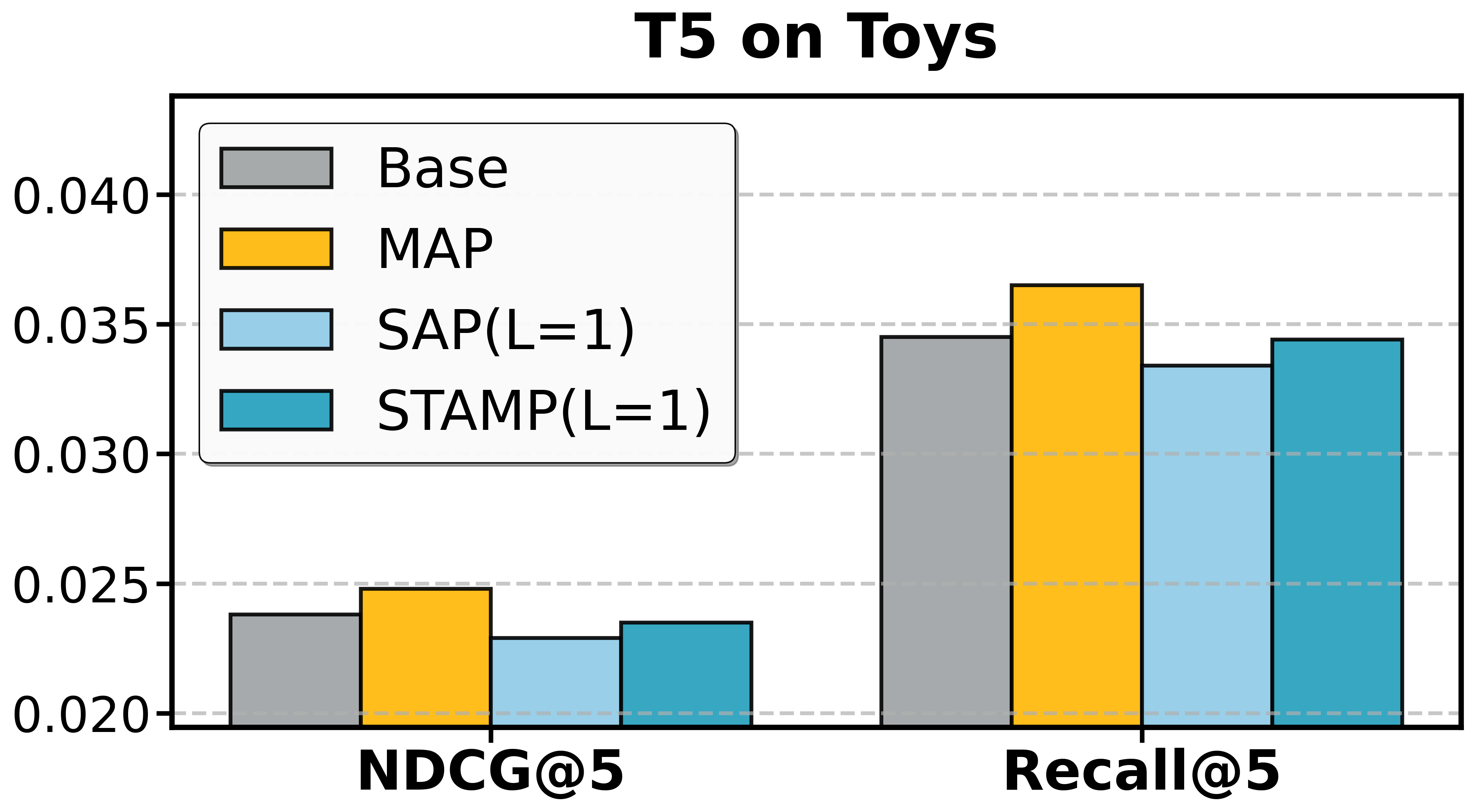}
    \end{subfigure}
    \hfill
    \begin{subfigure}[b]{0.495\linewidth}
        \includegraphics[width=\textwidth]{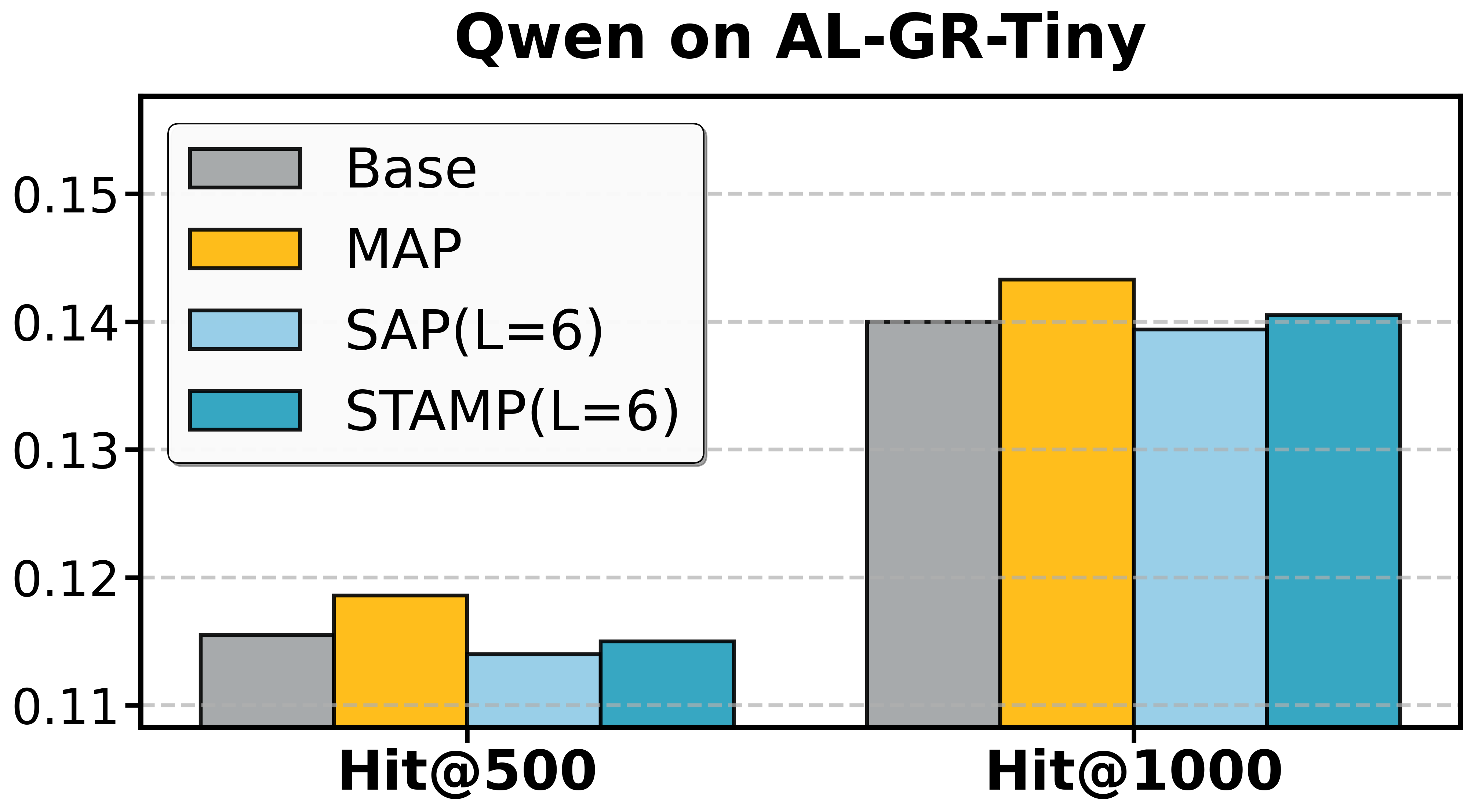}
    \end{subfigure}
    \vspace{-0.2cm}
    \caption{Ablation study of T5 on Amazon datasets and Qwen on AL-GR-Tiny.}
    \Description{Four grouped bar charts in reading order show T5 on Beauty, Sports, and Toys, then Qwen on AL-GR-Tiny. Horizontal categories are NDCG@5 and Recall@5 for T5, and Hit@500 and Hit@1000 for Qwen; vertical axes show scores on different, truncated scales. Gray, yellow, light blue, and teal denote Base, MAP, SAP, and STAMP, respectively, with pruning at layer 1 for T5 and layer 6 for Qwen. MAP exceeds Base throughout. STAMP exceeds SAP throughout; on Beauty it also exceeds Base, while on Sports and Toys it remains slightly below Base. On AL-GR-Tiny, STAMP is slightly below Base at Hit@500 and slightly above it at Hit@1000, with scores of 0.1405 and 0.1400, respectively.}
    \label{fig:ablation_all}
\end{figure}



\subsubsection{\textbf{Performance on FORGE (Qwen)}}
Table \ref{tab:forge_qwen_algr_tiny} shows dynamics broadly consistent with GRID, with nuances from structural and data differences.
\begin{itemize}[leftmargin=1.6em, labelsep=0.6em, itemsep=0.2em, topsep=2pt]
    \item \textbf{Efficiency:} STAMP ($L=6$) reduces VRAM by 17.2\%, more modest than on T5 because (i) we use a conservative $\alpha=0.6$ to preserve dense collaborative signals, and (ii) Qwen2.5-0.5B's 24 layers mean more static VRAM is occupied by model parameters rather than data-dependent activations. The relative saving thus depends on both the retained sequence length and the balance between parameter memory and token-dependent activations. Saving several gigabytes per GPU is useful for industrial-scale training constrained by GPU memory.
    \item \textbf{Effectiveness:} AL-GR-Tiny's multimodal, collaborative-signal-enriched SIDs~\cite{forge} are exceptionally information-dense, so EARN's bottleneck causes even sharper degradation here, and SAP-only pruning (Section~\ref{RQ2}) similarly loses accuracy to irrecoverable information loss. STAMP substantially narrows this gap: MAP's densified supervision recovers most of the lost accuracy, landing within 0.001 of base on Hit@20/100/500 and slightly exceeding it on Hit@1000.
\end{itemize}

\subsection{In-depth Analysis}
\subsubsection{Ablation Study (RQ2).}
\label{RQ2}
\begin{figure}[!t]
    \centering
    \begin{subfigure}[b]{0.495\linewidth}
        \centering
        \includegraphics[width=\textwidth]{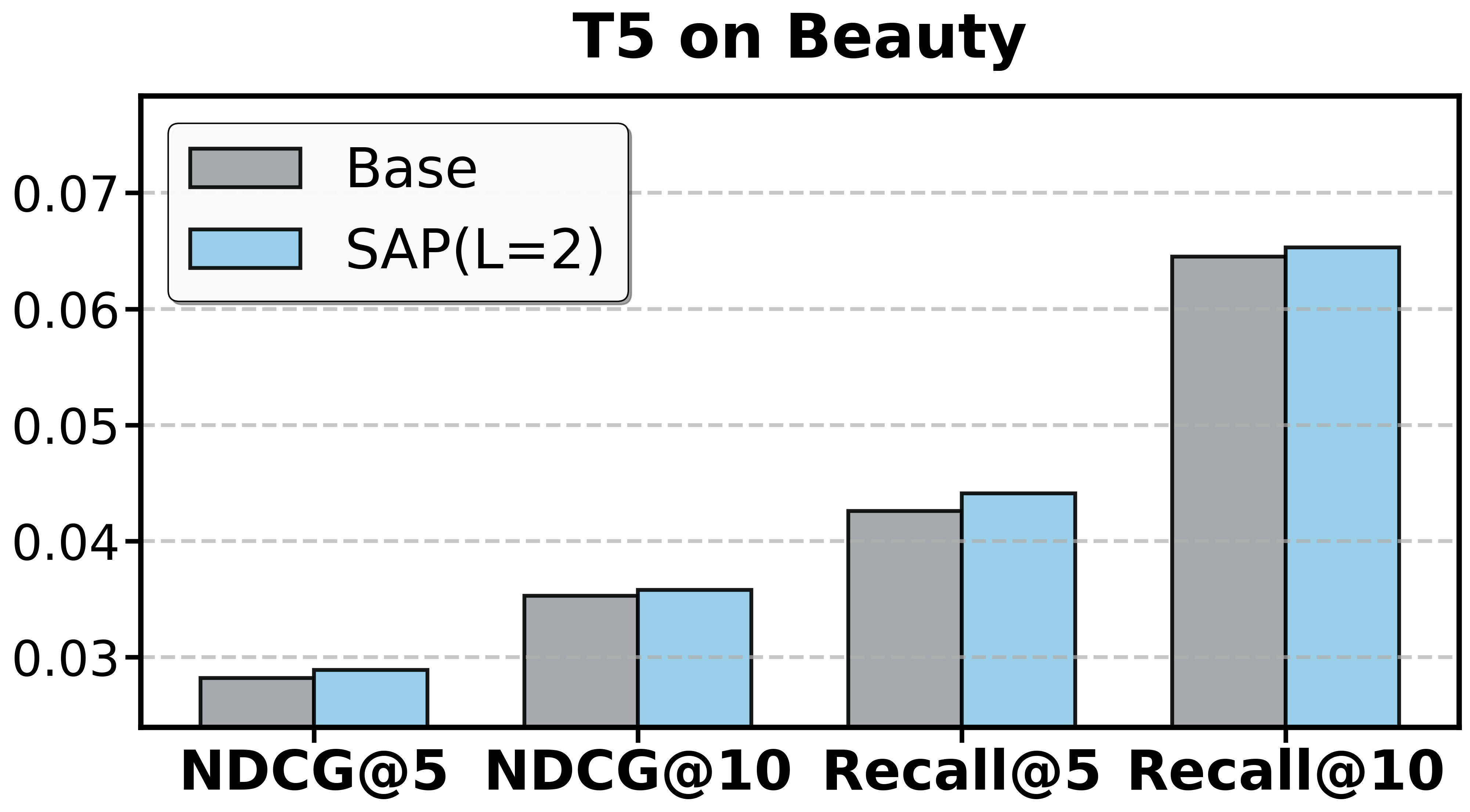}
    \end{subfigure}
    \hfill 
    \begin{subfigure}[b]{0.495\linewidth}
        \centering
        \includegraphics[width=\textwidth]{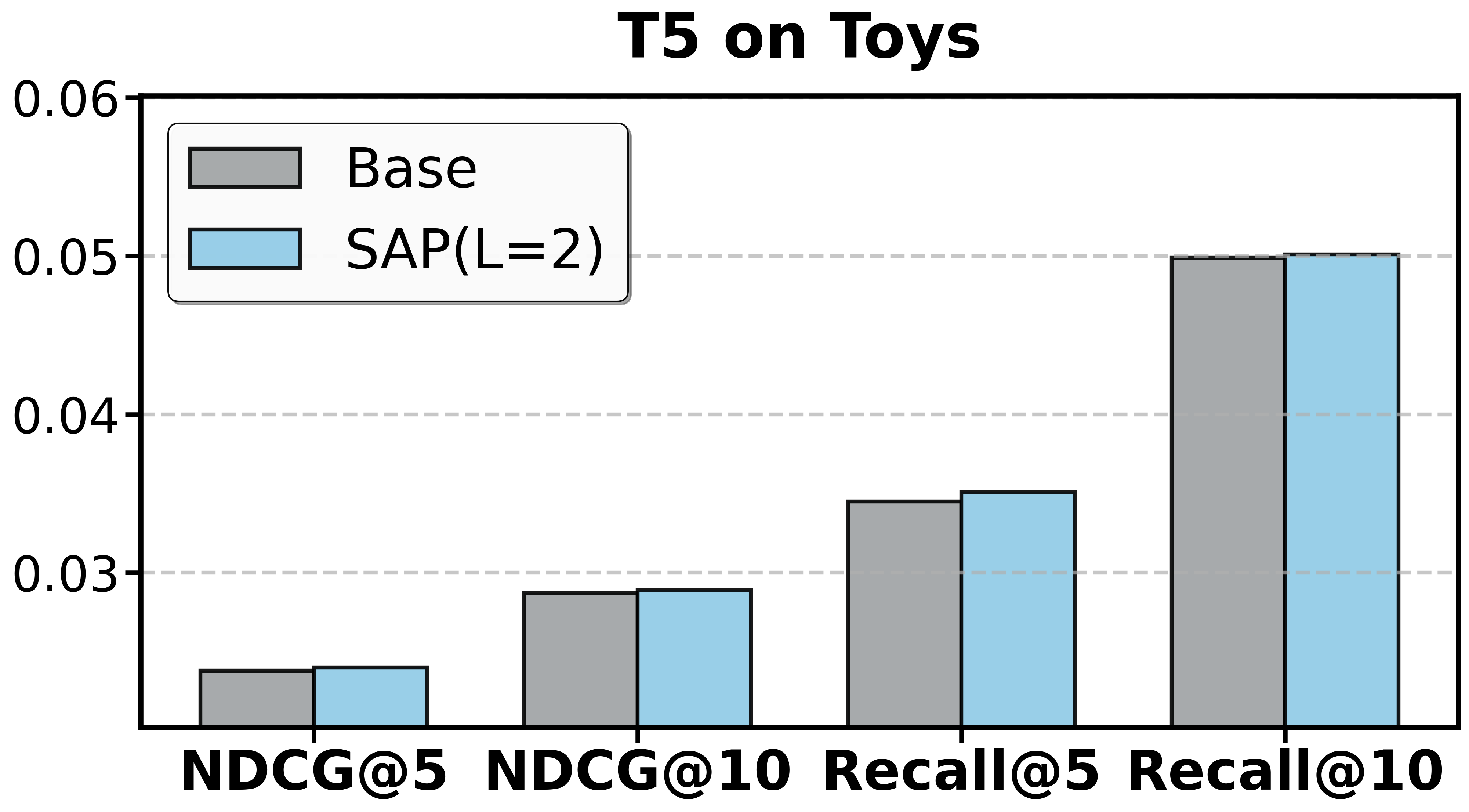}
    \end{subfigure}
    \vspace{-0.2cm}
    \caption{\textbf{Base vs. SAP ($L=2$) of T5 on Beauty and Toys.}}
    \Description{Side-by-side grouped bar charts for Beauty (left) and Toys (right) compare gray Base bars with light-blue SAP bars for pruning after layer 2. The horizontal axes list NDCG@5, NDCG@10, Recall@5, and Recall@10; vertical axes show metric values on dataset-specific, truncated scales. SAP is slightly higher than Base for each of the four metrics in both panels.}
    \label{fig:l2_comparison}
\end{figure}

To isolate the contributions of \textbf{SAP} and \textbf{MAP}, Figure~\ref{fig:ablation_all} reports accuracy (T5 on three Amazon datasets, Qwen on AL-GR-Tiny) and Table~\ref{tab:efficiency_final_styled} reports efficiency. Three observations emerge consistently:

\begin{table}[htbp]
    \centering
    \caption{\textbf{Efficiency Analysis.}}
    \label{tab:efficiency_final_styled}
    
    \definecolor{lightgray}{gray}{0.95}
    \setlength{\tabcolsep}{2.5pt}
    \renewcommand{\arraystretch}{1.1} 
    \resizebox{\linewidth}{!}{
    \begin{tabular}{|c|c|l|c|c|c|c|}
        \noalign{\hrule height 1.5pt}
        \textbf{Backbone} & \textbf{Dataset} & \textbf{Method} & \textbf{Time (s)} & \textbf{Speedup} & \textbf{VRAM (MiB)} & \textbf{Red.} \\
        \noalign{\hrule height 0.5pt}
        \noalign{\hrule height 0.5pt}
        
        \multirow{18}{*}{\textbf{T5}} & \multirow{6}{*}{Beauty} 
          & Base & 212 & - & 22234 & - \\ \cline{3-7}
        & & \cellcolor{lightgray}MAP & \cellcolor{lightgray}216 & \cellcolor{lightgray}0.98$\times$ & \cellcolor{lightgray}22238 & \cellcolor{lightgray}$\sim$0\% \\ \cline{3-7}
        & & SAP (L=2) & 165 & 1.28$\times$ & 13858 & 37.7\% \\ \cline{3-7}
        & & \cellcolor{lightgray}STAMP (L=2) & \cellcolor{lightgray}170 & \cellcolor{lightgray}1.25$\times$ & \cellcolor{lightgray}13862 & \cellcolor{lightgray}37.7\% \\ \cline{3-7}
        & & SAP (L=1) & 146 & 1.45$\times$ & 10374 & \textbf{53.3\%} \\ \cline{3-7}
        & & \cellcolor{lightgray}STAMP (L=1) & \cellcolor{lightgray}154 & \cellcolor{lightgray}1.38$\times$ & \cellcolor{lightgray}10378 & \cellcolor{lightgray}53.3\% \\ 
        
        \cline{2-7} 
        
        & \multirow{6}{*}{Sports} 
          & Base & 205 & - & 29370 & - \\ \cline{3-7}
        & & \cellcolor{lightgray}MAP & \cellcolor{lightgray}208 & \cellcolor{lightgray}0.99$\times$ & \cellcolor{lightgray}29374 & \cellcolor{lightgray}$\sim$0\% \\ \cline{3-7}
        & & SAP (L=2) & 160 & 1.28$\times$ & 18314 & 37.6\% \\ \cline{3-7}
        & & \cellcolor{lightgray}STAMP (L=2) & \cellcolor{lightgray}166 & \cellcolor{lightgray}1.23$\times$ & \cellcolor{lightgray}18318 & \cellcolor{lightgray}37.6\% \\ \cline{3-7}
        & & SAP (L=1) & 145 & 1.41$\times$ & 13298 & 54.7\% \\ \cline{3-7}
        & & \cellcolor{lightgray}STAMP (L=1) & \cellcolor{lightgray}151 & \cellcolor{lightgray}1.36$\times$ & \cellcolor{lightgray}13302 & \cellcolor{lightgray}54.7\% \\ 
        
        \cline{2-7}
        
        & \multirow{6}{*}{Toys} 
          & Base & 216 & - & 22268 & - \\ \cline{3-7}
        & & \cellcolor{lightgray}MAP & \cellcolor{lightgray}219 & \cellcolor{lightgray}0.99$\times$ & \cellcolor{lightgray}22272 & \cellcolor{lightgray}$\sim$0\% \\ \cline{3-7}
        & & SAP (L=2) & 169 & 1.28$\times$ & 14252 & 36.0\% \\ \cline{3-7}
        & & \cellcolor{lightgray}STAMP (L=2) & \cellcolor{lightgray}174 & \cellcolor{lightgray}1.24$\times$ & \cellcolor{lightgray}14256 & \cellcolor{lightgray}36.0\% \\ \cline{3-7}
        & & SAP (L=1) & 153 & 1.41$\times$ & 10654 & 52.2\% \\ \cline{3-7}
        & & \cellcolor{lightgray}STAMP (L=1) & \cellcolor{lightgray}159 & \cellcolor{lightgray}1.36$\times$ & \cellcolor{lightgray}10658 & \cellcolor{lightgray}52.1\% \\ 
        
        \hline 
        
        \multirow{4}{*}{\textbf{Qwen}} & \multirow{4}{*}{AL-GR-Tiny} 
          & Base & 1139 & - & 2*64232 & - \\ \cline{3-7}
        & & \cellcolor{lightgray}MAP & \cellcolor{lightgray}1152 & \cellcolor{lightgray}0.99$\times$ & \cellcolor{lightgray}2*65786 & \cellcolor{lightgray}-2.42\% \\ \cline{3-7}
        & & SAP (L=6) & 838 & 1.36$\times$ & 2*51632 & 19.6\% \\ \cline{3-7}
        & & \cellcolor{lightgray}STAMP (L=6)& \cellcolor{lightgray}851 & \cellcolor{lightgray}1.34$\times$ & \cellcolor{lightgray}2*53196 & \cellcolor{lightgray}17.2\% \\ \hline
        
        \noalign{\hrule height 1.5pt}
    \end{tabular}
    }
\end{table}
\begin{itemize}
    \item \textbf{MAP} alone consistently outperforms Base while retaining the full input sequence. Its gain therefore reflects the benefit of auxiliary supervision independently of SAP, consistent with the Supervision Sparsity premise. Predicting additional future tokens gives the backbone more opportunities to learn shared representations from the same behavioral trajectory, helping explain why the auxiliary objective can improve accuracy even without pruning. This overhead scales with hidden dimension (the auxiliary MLP operates on $\operatorname{Concat}(\mathbf{h}_t, E(y_{t+1}))$): negligible for T5 ($d=128$, $\sim$0\% VRAM change) but more visible for Qwen ($d=896$, +2.42\%), an acceptable trade-off given MAP is discarded at inference.
    \item \textbf{SAP} reduces latency and VRAM, usually with a small accuracy trade-off. On T5 \textit{Beauty} and \textit{Toys}, however, pruning after the second encoder layer improves accuracy over the base model (Figure~\ref{fig:l2_comparison}). One possible explanation, which we leave to future work to verify, is that once the initial layers have already aggregated information across the sequence, removing the remaining low-utility tokens serves as a denoising step: real-world interaction data is known to often contain label or interaction noise~\cite{northcutt2021pervasive}, and discarding these tokens would let the model focus its remaining capacity on more reliable signals.
    \item \textbf{STAMP} combines both modules' largely additive effects: each contributes independently rather than compensating for artifacts in the other. STAMP thus achieves SAP's efficiency and MAP's robustness together. The modules' effectiveness without tight coordination allows independent tuning for a given deployment scenario, making the combination useful for industrial settings requiring frequent, efficient retraining.
\end{itemize}

\begin{figure}
    \centering
    \includegraphics[width=0.98\linewidth]{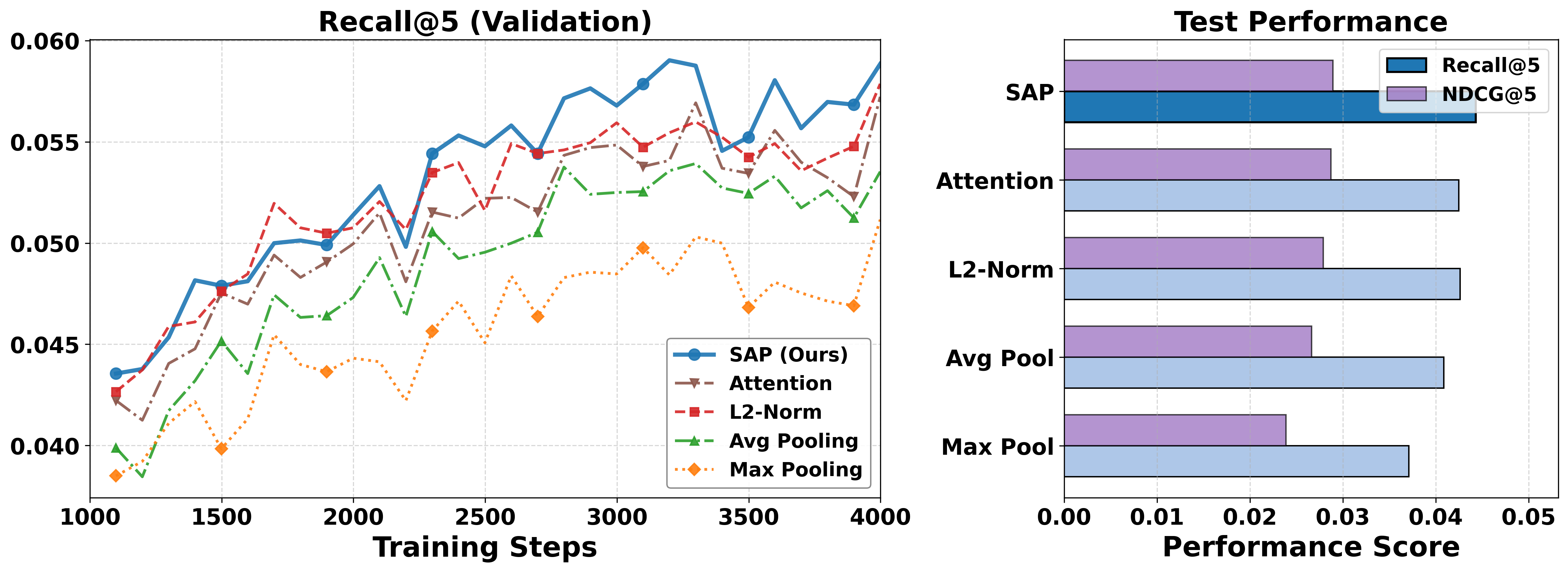}
    \vspace{-0.2cm}
    \caption{Comparison of Different Pruning Strategies (T5 on Beauty). (Left) Validation curves during the training phase. (Right) Performance on the test set.}
    \Description{Left: validation Recall@5 versus training steps from 1,000 to 4,000 compares SAP, attention-only, L2-norm, average pooling, and max pooling. All curves generally rise with fluctuations. Late in training, SAP is generally highest and max pooling generally lowest. Right: horizontal bars show test Recall@5 in blue and NDCG@5 in purple for the same methods. SAP has the highest scores on both metrics, attention-only and L2-norm are close behind, and the pooling methods are lower.}
    \label{fig:RQ3_1}
\end{figure}
\subsubsection{Impact of Pruning Strategies (RQ3)}
Within GRID, we compare SAP against (1) aggregation baselines (Max/Average Pooling) and (2) single-signal baselines using magnitude ($\ell_2$-norm) or attention centrality alone. Figure~\ref{fig:RQ3_1} shows training-phase validation curves (left) and final test results (right), yielding two insights. First, aggregation methods perform markedly worse: as seen in their lower training curves, blending token features together obscures the fine-grained semantic detail needed to capture precise user preferences, rather than merely discarding irrelevant tokens. Second, while the $\ell_2$-norm method performs reasonably by favoring large-magnitude tokens, it only inspects a token's internal value in isolation and ignores its contextual importance, so it may discard tokens with small individual magnitude but high incoming attention from others; the attention-only method has a complementary blind spot, favoring tokens that actively broadcast information to the rest of the sequence while overlooking the semantic aggregators that become important later by gradually absorbing information from other tokens. SAP consistently outperforms both single-signal baselines, with the gap widening in later training stages, confirming that accurately estimating token importance requires jointly considering semantic saliency and attention centrality rather than either alone.

\begin{figure}
    \centering
    \includegraphics[width=0.98\linewidth]{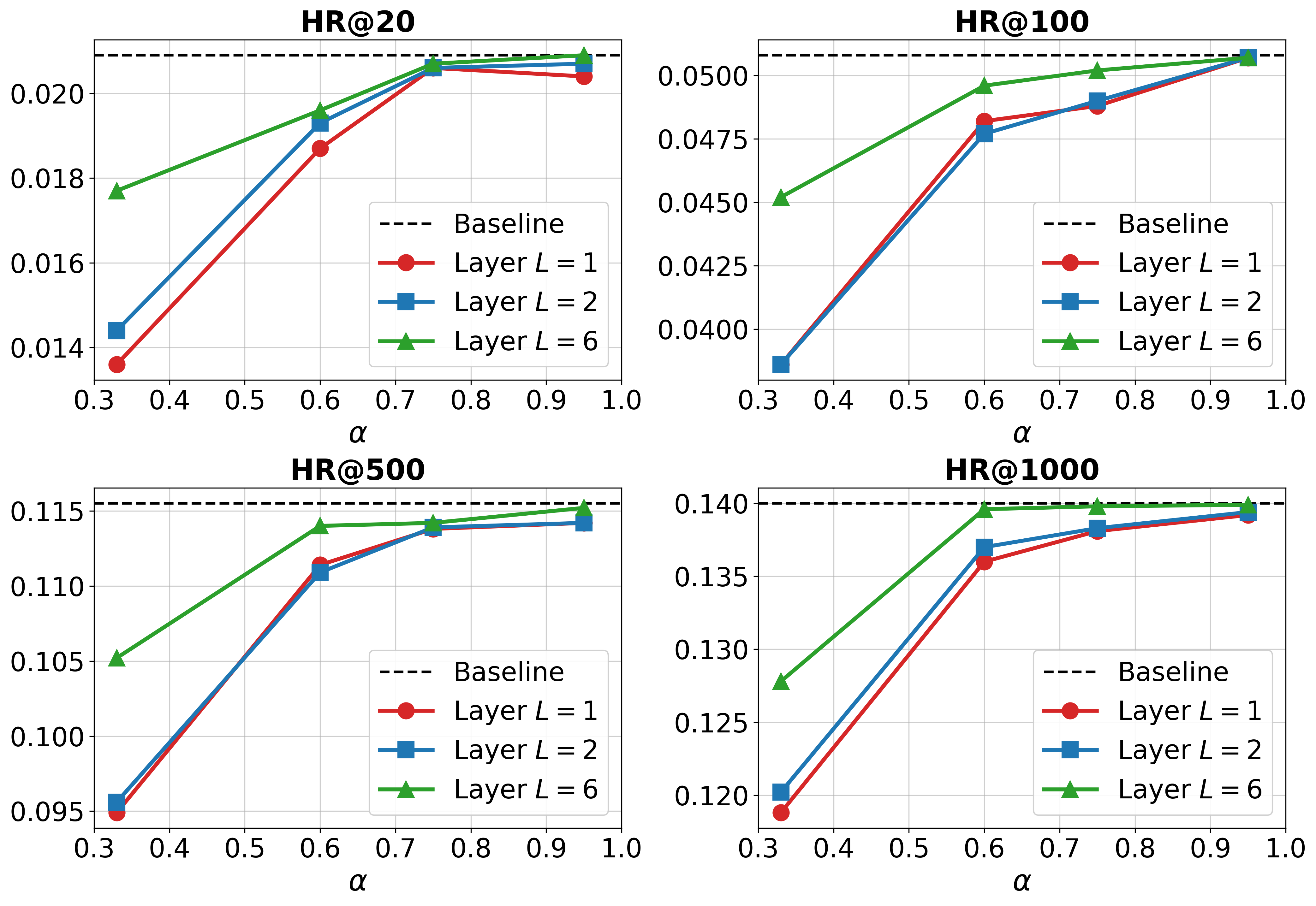}
    \vspace{-0.2cm}
    \caption{Performance of SAP with different Pruning Layer and Retention Ratio $\alpha$ (Qwen on AL-GR-Tiny).}
    \Description{Four line charts show HR@20, HR@100, HR@500, and HR@1000 in reading order, with retention ratio alpha on the horizontal axis and hit rate on the vertical axis. Red circles, blue squares, and green triangles denote pruning at layers 1, 2, and 6; black dashed lines mark the unpruned baseline. Hit rates generally approach the baseline as retention increases. Layer 6 is markedly better than layers 1 and 2 at the lowest retention setting, while differences between layers shrink at high retention.}
    \label{fig:RQ3_2}
\end{figure}
\begin{figure}
    \centering
    \includegraphics[width=0.98\linewidth]{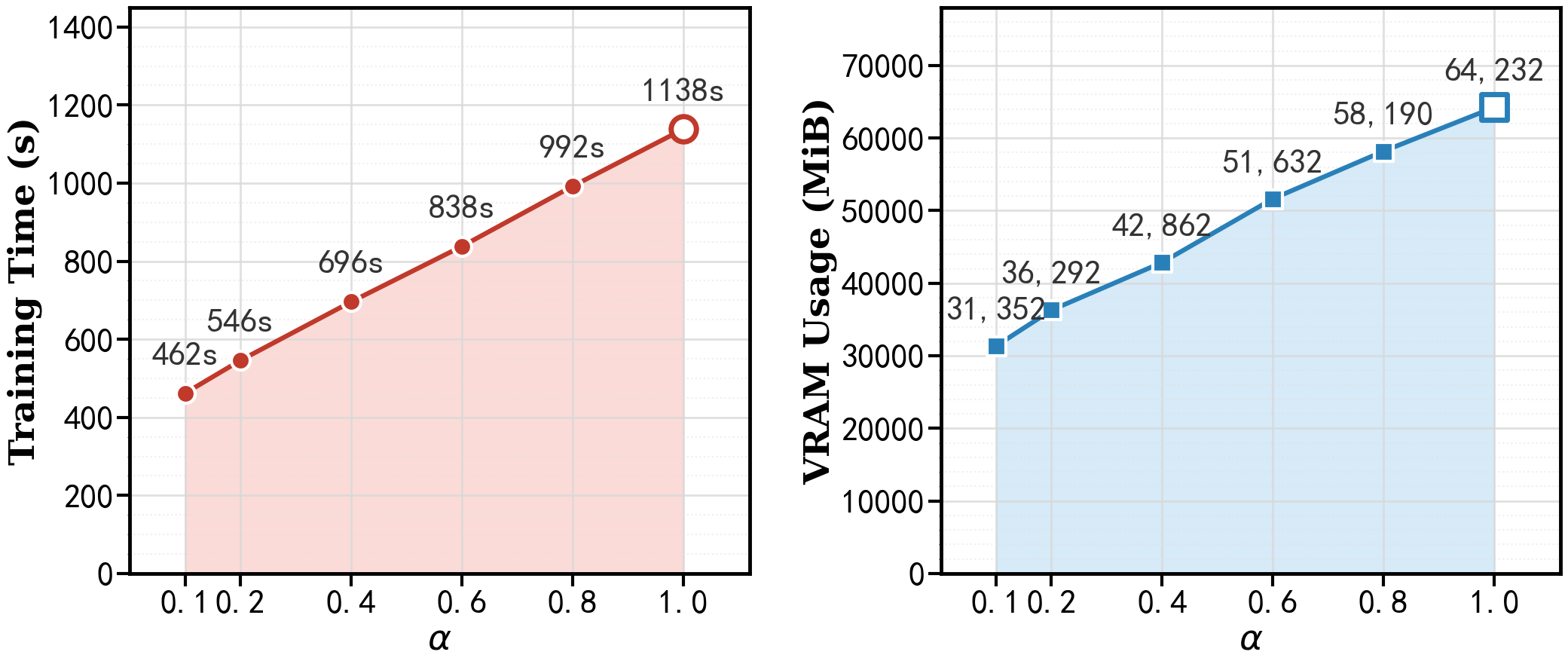}
    \vspace{-0.2cm}
    \caption{Efficiency of SAP (L=6) with different Retention Ratio $\alpha$ (Qwen on AL-GR-Tiny).}
    \Description{Two line charts share retention ratio alpha on the horizontal axis at 0.1, 0.2, 0.4, 0.6, 0.8, and 1.0. Left: red circles show training time in seconds, increasing from 462 to 1,138 as retention increases. Right: blue squares show VRAM usage in MiB, increasing from 31,352 to 64,232. Both costs rise monotonically; at alpha 0.6, the annotated values are 838 seconds and 51,632 MiB.}
    \label{fig:RQ3_3}
\end{figure}
\subsubsection{Hyperparameter Sensitivity: Pruning Layer and Retention Ratio (RQ3).}\label{RQ3.2}
Using Qwen2.5-0.5B-Instruct with SAP, we analyze the fundamental trade-off between training efficiency and downstream accuracy across the pruning layer ($L$) and retention ratio ($\alpha$). Earlier-layer pruning or lower retention ratios reduce forward token propagation, improving training speed and VRAM savings but increasing the risk of severe performance degradation. As shown in Figure~\ref{fig:RQ3_2}, the model exhibits remarkable tolerance for moderate compression, consistent with the Information Non-uniformity premise (Section~\ref{sec:preliminaries}) that a substantial proportion of input tokens is functionally redundant for the core recommendation task; however, predictive accuracy deteriorates sharply once compression surpasses a critical threshold, indicating that excessive pruning begins removing fine-grained semantic information indispensable for accurate user profiling. We further observe that the optimal pruning depth is intrinsically coupled with the retention ratio: the specific choice of pruning layer yields negligible impact under conservative settings, but delaying pruning to deeper layers becomes substantially more effective under aggressive compression. Consistent with prior work in inference acceleration~\cite{earn}, this suggests shallow layers are fundamentally crucial for encoding basic features and remain highly sensitive to premature information loss, whereas deeper layers are more robust to structural token removal, having already gathered sufficient context.

Figure~\ref{fig:RQ3_3} complements this accuracy analysis with the training cost of SAP at $L=6$. Both training time and VRAM decrease monotonically as $\alpha$ is reduced, from 1,138\,s and 64,232\,MiB at $\alpha=1.0$ to 462\,s and 31,352\,MiB at $\alpha=0.1$; the setting $\alpha=0.6$ corresponds to the SAP configuration in Table~\ref{tab:efficiency_final_styled} (838\,s, 51,632\,MiB). The savings are smaller than the reduction in retained tokens because the first six layers still process the full sequence, and the memory occupied by model parameters, gradients, and optimizer states does not depend on $\alpha$. Read together with Figure~\ref{fig:RQ3_2}, this suggests choosing the retention ratio just above the accuracy threshold: a moderate ratio such as $\alpha=0.6$ already reduces training time by about 26\% and VRAM by 19.6\%, whereas more aggressive pruning adds further savings at a growing risk to accuracy.

\begin{figure}
    \centering
    \includegraphics[width=0.98\linewidth]{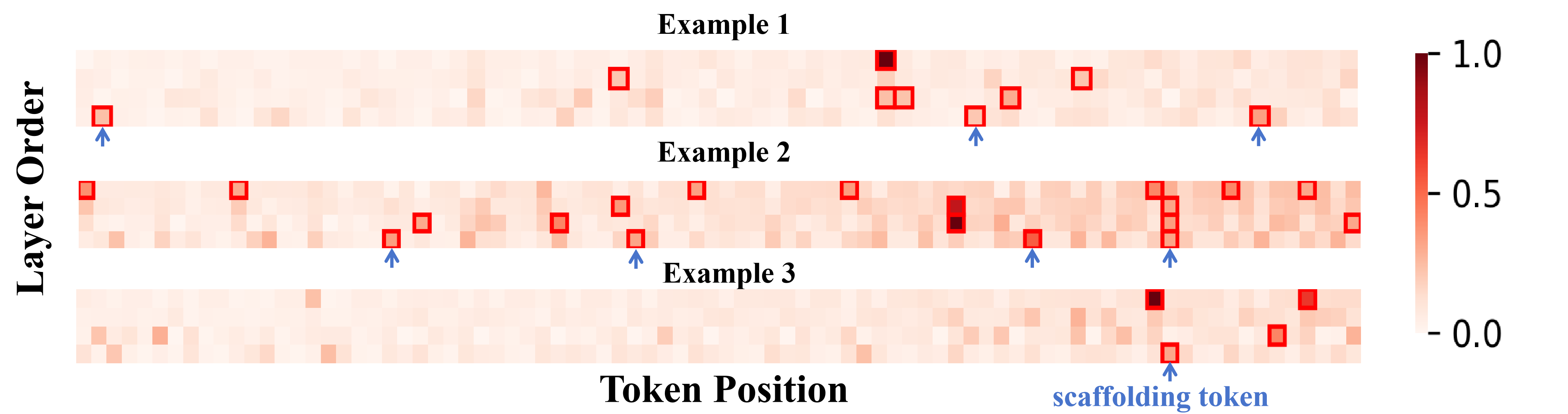}
    \caption{The attention distributions on different layers of three examples (T5 on Beauty).}
    \Description{Three stacked heatmaps, labeled Examples 1 to 3, place token position horizontally and four encoder layers vertically. A shared white-to-dark-red scale runs from 0 to 1 for attention intensity. Most cells are pale, with a few darker concentrations whose locations and strengths vary across layers and examples. Red outlines highlight selected cells, and blue arrows mark positions labeled scaffolding token. Several marked positions become paler across layers; the heatmaps show attention patterns rather than a direct test of whether removing those tokens preserves accuracy.}
    \label{fig:rq4}
\end{figure}
\subsubsection{Why is Pruning Safe? (RQ4)}
To understand why we can compress the majority of tokens without incurring performance loss, Figure~\ref{fig:rq4} visualizes T5 encoder attention heatmaps (Layers 1--4) for three randomly sampled sequences from the Beauty test set. Consistent with the information-aggregation phenomenon observed in LLMs~\cite{xiao2024efficient,guattention}, the attention patterns reveal a clear hierarchical information flow characterized by two key observations. First, a large portion of item tokens show consistently low attention scores throughout the entire encoding process; these likely represent noisy interactions or redundant information already overlapped by other dominant tokens. Their limited participation in user profiling suggests that pruning them can save computation. Second, a distinct subset of tokens exhibits a dynamic pattern, receiving high attention weights in early layers but fading rapidly in deeper ones. This suggests early layers use these tokens as scaffolding to build basic features, which are later absorbed by a smaller set of key tokens that then carry the aggregated information forward. Crucially, this explains why pruning should not occur too early: removing these scaffolding tokens before their information has been absorbed disrupts the initial aggregation process, whereas pruning them later, once their semantic value has already been digested by the remaining tokens, is both safe and efficient. 
These findings align with the earlier quantitative observation that shallow-layer pruning is substantially more sensitive than deep-layer pruning, and provide qualitative support for Information Non-uniformity.

\section{Conclusion}
We framed inefficiency and instability in SID-based Generative Recommendation through the Semantic Dilution Effect, a hypothesized shared source characterized by Information Non-uniformity at the input and Supervision Sparsity at the output. To address these interlinked challenges, we proposed STAMP, a dual-end framework reconciling training efficiency with model robustness. Its Semantic Adaptive Pruning (SAP) module dynamically filters redundant tokens during the forward pass, shortening the sequence processed by subsequent layers to reduce computational and memory overhead. Its Multi-step Auxiliary Prediction (MAP) module independently densifies supervision throughout the sequence, strengthening long-range dependency capture and enabling robust representation learning from sparse behavioral trajectories. MAP complements input compression rather than merely compensating for information loss. Its auxiliary branch is discarded at inference, so the additional supervision incurs no inference-time overhead. Extensive experiments across diverse datasets and backbone architectures (Encoder-Decoder and Decoder-Only) show that STAMP achieves significant training acceleration and memory savings while consistently maintaining, and in some cases surpassing, the recommendation accuracy of full-sequence baselines. An empirical analysis further reveals that the underlying attention mechanism naturally concentrates on a sparse subset of discriminative tokens, offering qualitative evidence consistent with Information Non-uniformity and suggesting why aggressive pruning can remain safe; a direct quantitative validation of the Semantic Dilution Effect is left to future work.

\begin{acks}
This work was supported by the National Natural Science Foundation of China (No. 62441236, 62402429, U24A20326), the Key Research and Development Program of Zhejiang Province (No. 2025C01026), the Ningbo Yongjiang Talent Introduction Programme (2023A-397-G), and the Young Elite Scientists Sponsorship Program by CAST (2024QNRC001).
\end{acks}

\clearpage
\section{Ethical Considerations}
All datasets used in this work are publicly available and have been anonymized prior to release, containing no personally identifiable information. No private user data was collected or used at any stage of this research. Therefore, we do not foresee significant ethical concerns regarding privacy, fairness, or potential misuse arising from this work.
\bibliographystyle{ACM-Reference-Format}
\balance
\bibliography{main}

\end{document}